\documentclass[%
 reprint,
 amsmath,amssymb,
 aps,showpacs
]{revtex4-2}

\usepackage{graphicx}
\usepackage{dcolumn}
\usepackage{bm}
\usepackage{float}
\usepackage{subcaption}
\usepackage{booktabs}
\usepackage{mathtools}
\usepackage{threeparttable}
\usepackage{caption}
\usepackage[inline]{trackchanges} 
\addeditor{KK}
\addeditor{NS}

\usepackage{soul} 

\definecolor{light-gray}{gray}{0.85}

\newcommand{\drawSlope}[7]{ 
				\coordinate (center1) at (#1,#2); 
				 \FPeval{\nx}{cos(#4*pi/180)}%
				 \FPeval{\ny}{sin(#4*pi/180)}%
				 \FPeval{\absNx}{abs(\nx)}%
				 \FPeval{\absNy}{abs(\ny)}%
				\coordinate (b) at ($ (center1) + #3*(-\ny,+\nx) $);
				\coordinate (c) at ($ (center1) - #3*(-\ny,+\nx) $); 
				 \FPeval{\xx}{(-\nx*\ny)}%
				\ifdim\xx pt < 0pt 
				\coordinate (d) at ($ (c) -2*#3*\ny*(1,0) $);
				\node at ($(d) +(0,#3*\nx)-0.2*(\nx/\absNx,0)$) {{\color{#5}$\scriptstyle #6$}}; 
				\node at ($(d)+(#3*\ny,0)-0.12*(0,\ny/\absNy)$) {{\color{#5}$\scriptstyle #7$}}; 
				\else
				\coordinate (d) at ($ (c) +2*#3*\nx*(0,1) $);
				\node at ($(d)-#3*\nx*(0,1)-0.2*\nx/\absNx*(1,0)$) {{\color{#5}$\scriptstyle #6$}}; 
				\node at ($(d)-#3*\ny*(1,0)-0.12*\ny/\absNy*(0,1)$) {{\color{#5}$\scriptstyle #7$}}; 
				\fi
				\draw[#5,line width=0.1mm] (d) -- (b); 
				\draw[#5,line width=0.1mm] (d) -- (c); 
				\draw[#5,line width=0.1mm] (b) -- (c); 
} 

\usepackage{graphicx}
\usepackage{amsmath}
\usepackage{mathtools} 
\usepackage{color}
\usepackage{overpic}
\usepackage[export]{adjustbox}
\usepackage{mwe, tikz}
\usepackage{rotating}
\usepackage{latexsym}
\usepackage{csquotes}
\usepackage{tikz} 
\usetikzlibrary{calc,arrows,decorations.markings} 
\usepackage{pict2e}
\usepackage{pgfplots}
\usepackage{float}
\usepackage{hyperref}
\usepackage{xr}
\usepackage{rotating}
\usepackage{amsmath,amssymb}
\usepackage{fp}
\usetikzlibrary{calc} 
\usepackage{booktabs, tabularx}
\usepackage{romannum}

\soulregister\ref7
\soulregister\cite7
\soulregister\citeA7
\soulregister\romannum7

\begin{document}

\preprint{APS/123-QED}

\title{Sheared granular matter \& the empirical relations of seismicity}
\author{Nauman Hafeez Sultan}
\author{Kamran Karimi}%
\email{email address: kamran.karimi1@ucalgary.ca}
\author{J\"orn Davidsen}
\altaffiliation[Also at:]{
Hotchkiss Brain Institute, University of Calgary, 3330 Hospital Dr NW, Calgary, Alberta T2N 4N1, Canada}
\affiliation{
Complexity Science Group, Department of Physics and Astronomy, University of Calgary, 2500 University Drive NW, Calgary, Alberta, Canada T2N 1N4
}%


\begin{abstract}
The frictional instability associated with earthquake initiation and earthquake dynamics is believed to be mainly controlled by the dynamics of fragmented rocks within the fault gauge. Principal features of the emerging seismicity (e.g. intermittent dynamics and broad time and/or energy scales) have been replicated by simple experimental setups, which involve a slowly driven slider on top of granular matter, for example. Yet, these set-ups are often physically limited and might not allow one to determine the underlying nature of specific features and, hence, the universality and generality of the experimental observations. Here, we address this challenge by a numerical study of a spring-slider experiment based on two dimensional discrete element method simulations, which allows us to control the properties of the granular matter and of the surface of the slider, for example. Upon quasi-static loading, stick-slip-type behavior emerges which is contrasted by a stable sliding regime at finite driving rates, in agreement with experimental observations. Across large parameter ranges for damping, inter-particle friction, particle polydispersity etc. the earthquake-like dynamics associated with the former regime results in several robust scale-free statistical features also observed in experiments. At first sight these closely resemble the main empirical relations of tectonic seismicity at geological scales. This includes the Gutenberg-Richter distribution of event sizes, the Omori-Utsu-type decay of aftershock rates, as well as the aftershock productivity relation and broad recurrence time distributions. Yet, we show that the correlations associated with tectonic aftershocks are absent such that the origin of the Omori-Utsu relation, the aftershock productivity relation, and B{\aa}th's relation in the simulations is fundamentally different from the case of tectonic seismicity. We argue that the same is true for previous lab experiments.

\end{abstract}

\maketitle


\section{\label{sec:Intro}Introduction}
The frictional instability is a commonly observed phenomenon in a wide class of physical settings ranging from plastically deforming solids \cite{bardet1990comprehensive,karimi2018correlation,karimi2019plastic} and fractured rocks \cite{jaeger2009fundamentals,tal2020experimental,kwiatek2014seismic,goebel2014seismic,w2013acoustic,goebel2015comparison}  in laboratory-based experiments to faulting and landslides at geological scales \cite{scholz2002mechanics}. 
Under a slow driving rate, this mechanism leads to an \emph{irrecoverable} slip motion that a stuck system undergoes beyond its frictional threshold, hence the term ``stick-slip" instability \cite{rabinowicz1958intrinsic}.
The emergent dynamics exhibits highly intermittent features, the so-called ``avalanches", with a broad range of associated time, length, and energy scales.
The scale-free nature of avalanche statistics may be considered as a signature of a dynamical \emph{yielding transition} which is characterized by universal scaling features such as diverging length and/or timescales and power-law distributions of avalanche sizes \cite{LinPNAS2014,karimi2017inertia}.

Essential features of this critical dynamics have been recovered in several numerical frameworks as well as experimental settings (see \cite{de2016statistical} and references therein). 
Recently, stick-slip behavior has been investigated in laboratory experiments using a simple spring-slider setup, choosing granular matter as a substrate undergoing plastic deformation \cite{zadeh2019seismicity,zadeh2019crackling,cheng2101correlating}.
This is thought to be a good candidate to study slip planes of fragmented rocks, for example.
The setup was reported to reveal critical avalanche dynamics under quasi-static loading conditions which transitioned through a breakdown of scaling features to a \emph{non}-critical regime of stable sliding regime at finite driving rates.
Within the former regime, essential (empirical) features and statistical relations of tectonic seismicity were recovered at the lab scale suggesting a (potentially) common physical mechanism across scales.
Observations of some of these empirical relations established for tectonic seismicity have also been reported for other laboratory experiments, including a shearing granular experiment with a cylindrical geometry~\cite{lherminier2019continuously}, dislocation and slip avalanches in (poly)crystals \cite{zaiser2006scale,weiss2003three} as well as macro fracturing in brittle rocks and other heterogeneous solids \cite{baro2013statistical,baro2018experimental,davidsen2017triggering}.

While such empirical observations led to the development of commonly-used constitutive models (such as the rate and state friction relation or viscoelastic rheology \cite{scholz2002mechanics}), we add the caveat that experimental set-ups are often limited by physical constraints such that extracting some of the vital information about the dynamics of the system has remained challenging.
Essential control parameters including (but not limited to) internal dissipation mechanisms~\cite{brilliantov1996model}, inter-grain friction~\cite{santos2020granular}, polydispersity~\cite{ma2020size}, particle shape~\cite{murphy2019transforming}, and the surface roughness of the slider are not easily tunable in real experimental conditions.
This motivates an interest in the development of numerical models to not only reproduce the observed experimental results but also to circumvent the experimental limitations in order to achieve a better understanding of the essential and controlling ingredients of the underlying physics.
At the same time, such frameworks have the potential to be useful across scales and in particular to allow relevant numerical measurements that would otherwise be impossible from direct seismological observations. In addition, they might help augment existing (often phenomenological) constitutive equations by incorporating micromechanical aspects of the deformation and failure of granular solids~\cite{scholz2002mechanics,de2016statistical}.
This, in turn, might improve their predictive power, which ultimately might lead to better seismic hazard assessment. 

This study develops a numerical model of the aforementioned spring-slider experiment on a granular substrate~\cite{zadeh2019seismicity,zadeh2019crackling,cheng2101correlating} using discrete element modelling (DEM) \cite{cundall1979discrete} to achieve the aforementioned goals. Our findings based on this model include previous experimental observations of empirical relations from tectonic seismicity such as the Gutenberg-Richter distribution of earthquakes magnitudes, an Omori-Utsu-like decay of seismic activity following large events and other established relations \cite{kagan2013earthquakes}. In this context, our numerical study allows us to address the origin of these relations and illuminate differences between previous lab experiments and tectonic seismicity.
In particular, we show that the statistical relations describing the dynamics of "aftershocks" have a different origin compared to tectonic seismicity. In fact, we find an absence of pronounced temporal correlations and clustering, which are one of the defining properties of aftershocks in tectonic settings, such that the notion of aftershocks in the spring-slider set-up becomes highly questionable.

The organization of the paper is as follows:
In Sec. \ref{sec:Model}, the shear setup, packing preparation, driving protocol, and relevant simulation details are discussed.
Section~\ref{sec:Transition} discusses the rate effects resulting in the transition between the stable sliding and stick-slip regimes.
In Sec. \ref{sec:Seimicity}, we quantify the statistics of avalanches including their size distribution, duration, as well as their temporal evolution. Sec.~\ref{sec:discussions} and Sec.~\ref{sec:conclusions} present discussions and conclusions, respectively.

\section{\label{sec:Model}Numerical Modelling}

\subsection{Slider-substrate setup}

\begin{figure}[b]
\begin{center}
\includegraphics[width=0.5\textwidth]{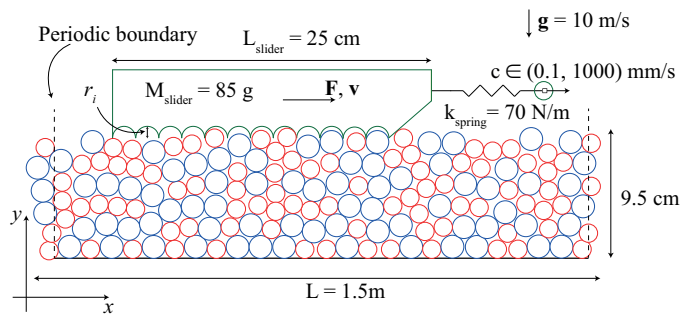}
\end{center}
\caption{\label{fig:SliderSystem} Granular simulation set-up (drawing not to scale). The discs represent the bulk sample with length $L$ and periodicity along $x$.}
\end{figure}

Our slider-substrate setup is created in accordance with the experiment detailed in \cite{zadeh2019crackling}. In our two-dimensional simulation the substrate granular disks are sheared using a (rigid) slider of length $L_{\text{slider}} = 25~cm$ and mass $M_\text{slider} = 8.5\times 10^{-2}~kg$. A spring of stiffness $k_\text{spring} = 70~N/m$ is attached to the slider, the free end of the spring is pulled at constant speed $c$. To minimize the rotation of the slider due to spring vibrations and uniformly distribute the force due to pulling along the entire length, the spring is anchored to the middle of the slider. A schematic drawing of this setup is shown in Fig.~\ref{fig:SliderSystem}.

To model this setup numerically Discrete Elements (DE) approach in LAMMPS \cite{plimpton1995fast} is chosen which considers each particle in the simulation as an individual element with a constant mass and radius. Furthermore, elements of this setup are fixed along the $z$-axis hence behaves as two dimensional. Substrate discs, total of $N = 7770$, are distributed bi-dispersedly with size ratio of $R_b/R_s = 1.25$ and number ratio $N_b/N_s = 2.5$. 
Here $N_{b(s)}$ and $R_{b(s)}$ denote the number of discs and their radii, respectively, we set $R_b = 2.5~mm$ for the larger disks.
The density of substrate particles is constant at $\rho = 2.5 \times 10^4 ~kg/m^3$, which defines the mass of the particle $m$. Gravity, acceleration of $10~m/s^2$, is acting downwards on the entire system.
The interaction between particles is modelled to have normal, tangential, and rolling forces, input parameters of these interactions are detailed in the next subsection.

As shown in Fig.~\ref{fig:SliderSystem}, the rigid slider for this simulation has circular grooves with radius $r_{i} = 2.5~mm$ on the base to induce additional friction.
To setup the substrate, particles are initially dropped under gravity in the simulation box of length, $L = 1.5$ m, filled up to height of $0.3L$, which has periodic boundaries along $x$. 
The slider is dropped on substrate particles after the (scaled) kinetic energy of the system is lower than $K=\sum_{i=1}^{N} m_i\mathbf{v}_i.\mathbf{v}_j/2N< 10^{-10}~J$.
The spring is attached once the slider comes to full stop above the substrate.
The whole system is allowed to relax, such that particles and slider are completely at rest before shearing begins.

\subsection{Mechanics of DE}

\begin{figure}[t]
\begin{center}
\includegraphics{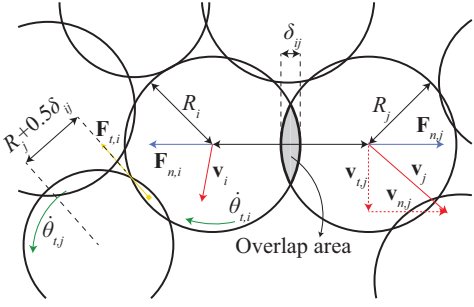}
\end{center}
\caption{\label{fig:forces} Interaction forces due to contact of particles (drawing not to scale).}
\end{figure}

The substrate disks are modelled as granular particles, for which the forces are updated at each timestep based on their interaction with one another, the slider and the walls. The parameters defined for interacting forces within the substrate and between the slider and the granular disks, discussed throughout this subsection, are outlined in Table~\ref{tab:Parameters}.

The $i$-th and $j$-th particles in Fig.~\ref{fig:forces} with position vectors $\mathbf{r}_i$ and $\mathbf{r}_j$ only interact when the overlap $\delta_{ij} = R_i+R_j - \left|\mathbf{r}_{ij}\right| >0$ with $\mathbf{r}_{ij} = \mathbf{r}_{i}-\mathbf{r}_{j}$.
The normal contact force is modelled as Hertz interaction, with the stiffness $k_n$, and is calculated for each timestep as
\begin{equation}\label{EqForceNormal}
\mathbf{F}_{n} = -k_{n}R^{1/2}_{\text{eff}}\delta^{3/2}_{ij}\mathbf{n},
\end{equation}
with the effective radius $R_\text{eff} = {R_{i}R_{j}}/(R_{i}+R_{j})$ and the unit normal vector $\mathbf{n} = {\mathbf{r}_{ij}}/{\left|\mathbf{r}_{ij}\right|}$.

The tangential force $\mathbf{F}_{t}$ between substrate particles is given by
\begin{equation}
\mathbf{F}_{t} = -\min (~\mu_{t}~|\mathbf{F}_{n}|,
~|-k_{t}\boldsymbol{\xi}_{t}+\mathbf{F}_{t}^{\text{damp}}|)~\mathbf{t},
\end{equation}
where $\mu_{t}$ is the tangential friction coefficient, $k_{t}$ is tangential stiffness of particles, $\boldsymbol{\xi}_{t} = \int_{T_{\text{cont}}} \mathbf{v}_{t}^{\text{rel}}(\tau)~d\tau$ is the tangential displacement accumulated during the entire duration of the contact $T_{\text{cont}}$, and $\mathbf{t} = {\mathbf{v}_{t}^{\text{rel}}}/{\left|\mathbf{v}_{t}^{\text{rel}}\right|}$ is the tangential unit vector.

Damping is evaluated using a viscoelastic damping model in LAMMPS, the normal (tangential) component of which is calculated by
\begin{equation}\label{EqDamping}
\mathbf{F}_{n(t)}^{\text{damp}} = -\eta_{n(t)}\mathbf{v}_{n(t)}^{\text{rel}}.
\end{equation}
The normal damping coefficient $\eta_{n}$ is given by
\begin{equation}
\eta_{n} = a~\eta_{n0}~m_{\text{eff}},
\end{equation}
with the damping prefactor $\eta_{n0}$, the radius of the contact area $a = \sqrt{R_{\text{eff}}~\delta_{ij}}$, and the effective mass $m_{\text{eff}} = {m_{ij}}/{(m_{i}+m_{j})}$.
We have $\eta_{t} \propto \eta_{n}$ with the proportionality factor $x_{\gamma t}$.
The normal relative velocity vector is $\mathbf{v}_{n}^{\text{rel}} = (\mathbf{v}_\text{rel}\cdot\mathbf{n})~\mathbf{n}$ where $\mathbf{v}_\text{rel}=\mathbf{v}_i-\mathbf{v}_j$. 
We define the relative tangential velocity $\mathbf{v}_{t}^{\text{rel}}$ as
\begin{equation}
\mathbf{v}_{t}^{\text{rel}} = \mathbf{v}_{t} - \left(R_{i}\dot\theta_{i} + R_{j}\dot\theta_{j}\right)\mathbf{e}_z\times \mathbf{n},
\end{equation}
with the angular velocity $\dot\theta_{i}\mathbf{e_z}$ of the $i$-th particle.


Rolling force during a contact is
\begin{equation}
\mathbf{F}_{r} = -\min (~\mu_{r}|\mathbf{F}_{n}|,
\left|-k_{r}\boldsymbol{\xi}_{r}-\gamma_{r}\mathbf{v}_{r}\right|)~\mathbf{k},
\end{equation}
where $\mu_r$ is the rolling friction, $k_{r}$ is rolling stiffness, $\gamma_{r}$ is the rolling dampness, $\mathbf{v}_{r} = R_{\text{eff}}\left(\dot\theta_{i}-\dot\theta_j\right)\mathbf{e}_z\times\mathbf{n}$ is the relative rolling velocity, $\boldsymbol{\xi}_{r} = \int_{T} \mathbf{v}_{r}(\tau)d\tau$ is rolling displacement, and $\mathbf{k}=\mathbf{v}_{r}/|\mathbf{v}_{r}|$.

\begin{table}[b]
\begin{ruledtabular}
\begin{threeparttable}
\begin{tabular}{lllll}
\textrm{Parameter}&
\textrm{Symbol}&
\textrm{Type \emph{i}\tnote{a}}& 
\textrm{Type \emph{ii}\tnote{b}}&  
\textrm{Units}\\
\colrule
Normal stiffness & $k_{n}$ & $1\times10^{6}$ & $1\times10^{6}$ & ${N/m}^2$\\
Normal damping & $\eta_{n0}$ & $5\times 10^{5}$ & $5\times 10^{5}$ & $1/{m}\cdot \text{s}$\\
Tangential stiffness & $k_{t}$ & $1\times10^{1}$ & $1\times10^{0}$ & ${N/m}$\\
Sliding friction & $\mu_{t}$ & $1\times10^{1}$ & $1\times10^{0}$ & -\\
Damping ratio & $x_{\gamma t}$ & $1\times10^{1}$ & $1\times10^{4}$ & -\\
Rolling friction & $\mu_{r}$ & $1\times10^{1}$ & $0$ & -\\
Rolling stiffness & $k_{r}$ & $1\times10^{1}$ & $0$ & ${N/m}$\\
Rolling damping & $\gamma_{r}$ & $1\times10^{4}$ & $0$ & $kg/s$
\end{tabular}
\begin{tablenotes}
     \item[a] Substrate/Substrate
     \item[b] Slider/Substrate
   \end{tablenotes}
\end{threeparttable}
\caption{Microscopic material parameters used for the granular model.} \label{tab:Parameters}
\end{ruledtabular}
\end{table}

Newton's equations of motion, updated every timestep, are
\begin{eqnarray}
m_i\ddot{\mathbf{r}}_i &=& \mathbf{F}_n+\mathbf{F}_n^\text{damp}+\mathbf{F}_{t}+m_i\mathbf{g}, \nonumber \\
m_i R^2_i\ddot\theta_i \mathbf{e}_z &=& R_i~\mathbf{n}\times(\mathbf{F}_t+\mathbf{F}_r).
\end{eqnarray}

The rate unit (inverse timescale) is set by gravity $\sqrt{{g}/{R_s}}$.
The normal vibrational frequency is defined as $\omega_{n}=\sqrt{k^\text{eff}_{n}/m^{\text{eff}}}$ with $k^\text{eff}_{n} = a k_n$.
For the example of normal force, our choice of microscopic parameters obeys the following separation of timescales
\begin{equation}
\omega_n \gg \sqrt{\frac{g}{R_{s}}} \gg \frac{c}{R_{s}}.
\end{equation}
This leads to an internal dynamics reasonably close to the experimental setting. 
We also set the discretization time $\Delta t=0.05~\omega_n^{-1}$.
Furthermore, the dissipation rate $\tau^{-1}_d = a\eta_{n0}$, of our simulations relates to the vibrations frequency as $\tau^{-1}_d/\omega_n\simeq 0.2$, to recover damped dynamics where the damping rates are such that they reproduce experimental observations. 



\newcommand{\ratio}{0.4}
\begin{figure}[t]
     \centering
     \begin{subfigure}[b]{\ratio\textwidth}
         \centering
         \includegraphics[width=\textwidth]{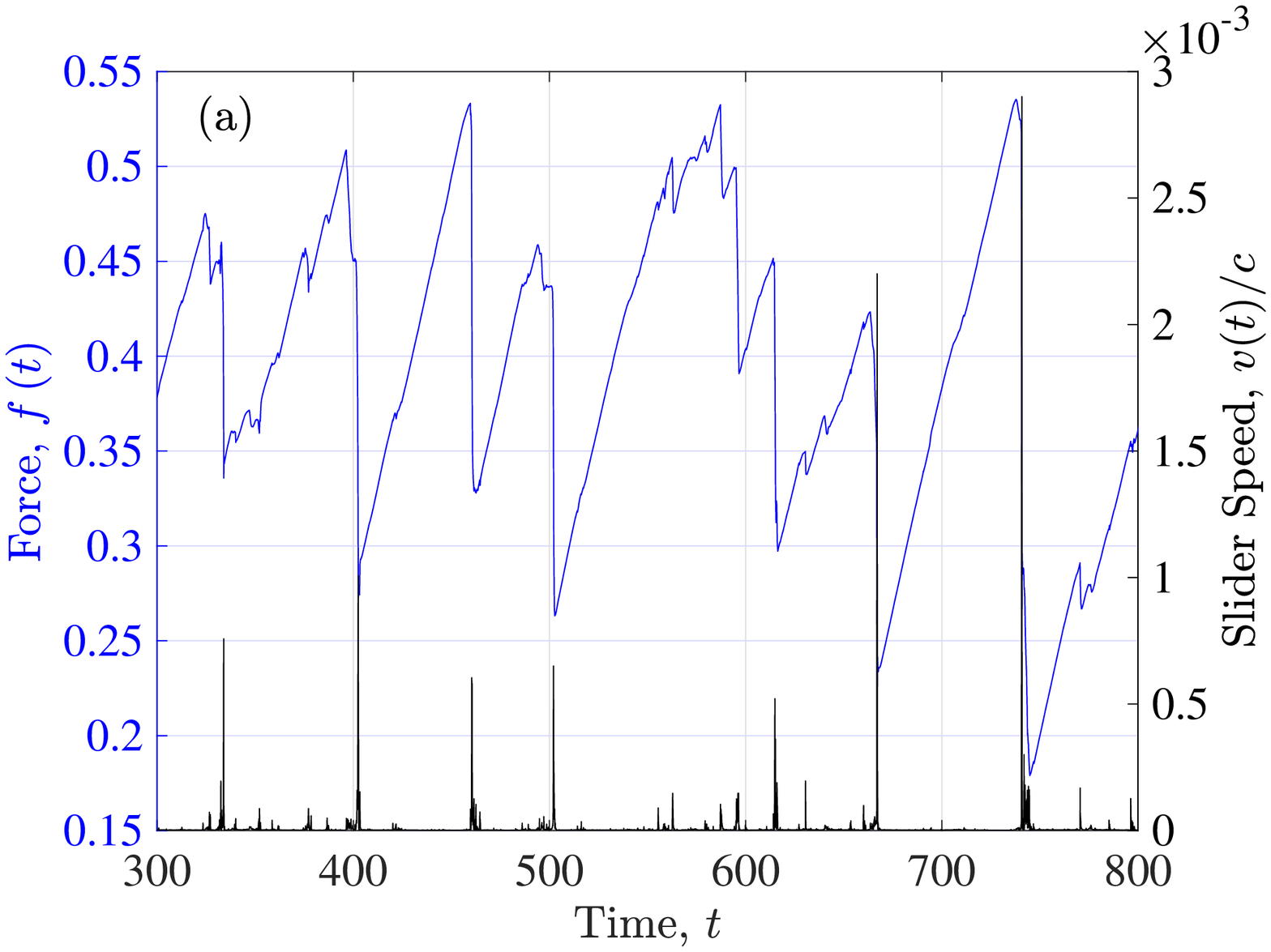}
         \label{fig:ForceC0.1}
     \end{subfigure}
     \hfill
     \begin{subfigure}[b]{\ratio\textwidth}
         \centering
         \includegraphics[width=\textwidth]{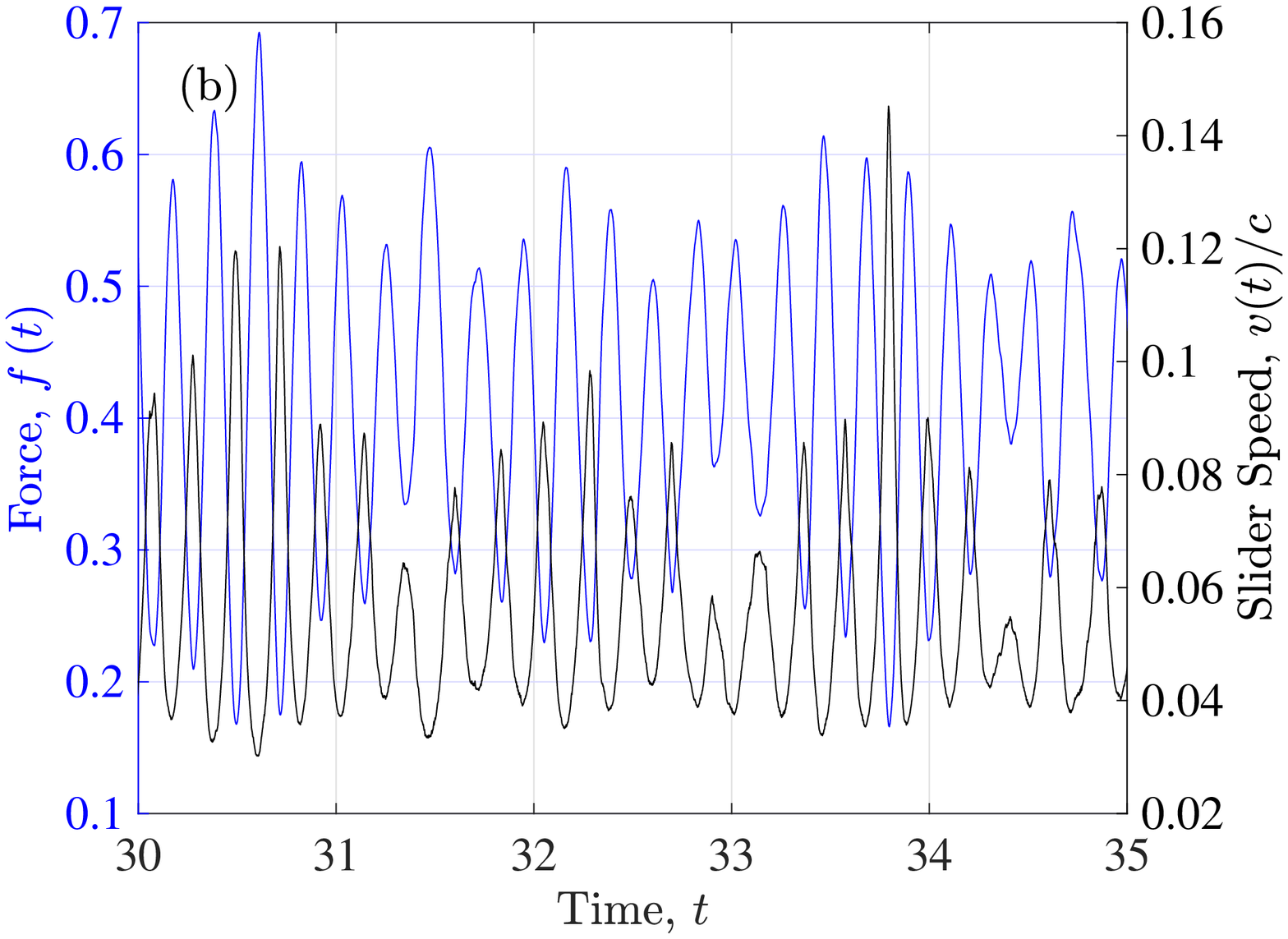}
         \label{fig:ForceC100}
     \end{subfigure}
        \caption{Force response $f(t)$ and slider speed $v(t)$ (normalized to the pulling speed $c$) plotted against $t$ at \textbf{a)} $c=0.1$ mm/s \textbf{b)} $c=1000$ mm/s.}
        \label{fig:Transition}
\end{figure}

\begin{figure}[t]
     \centering
     \begin{overpic}[width=\ratio\textwidth]{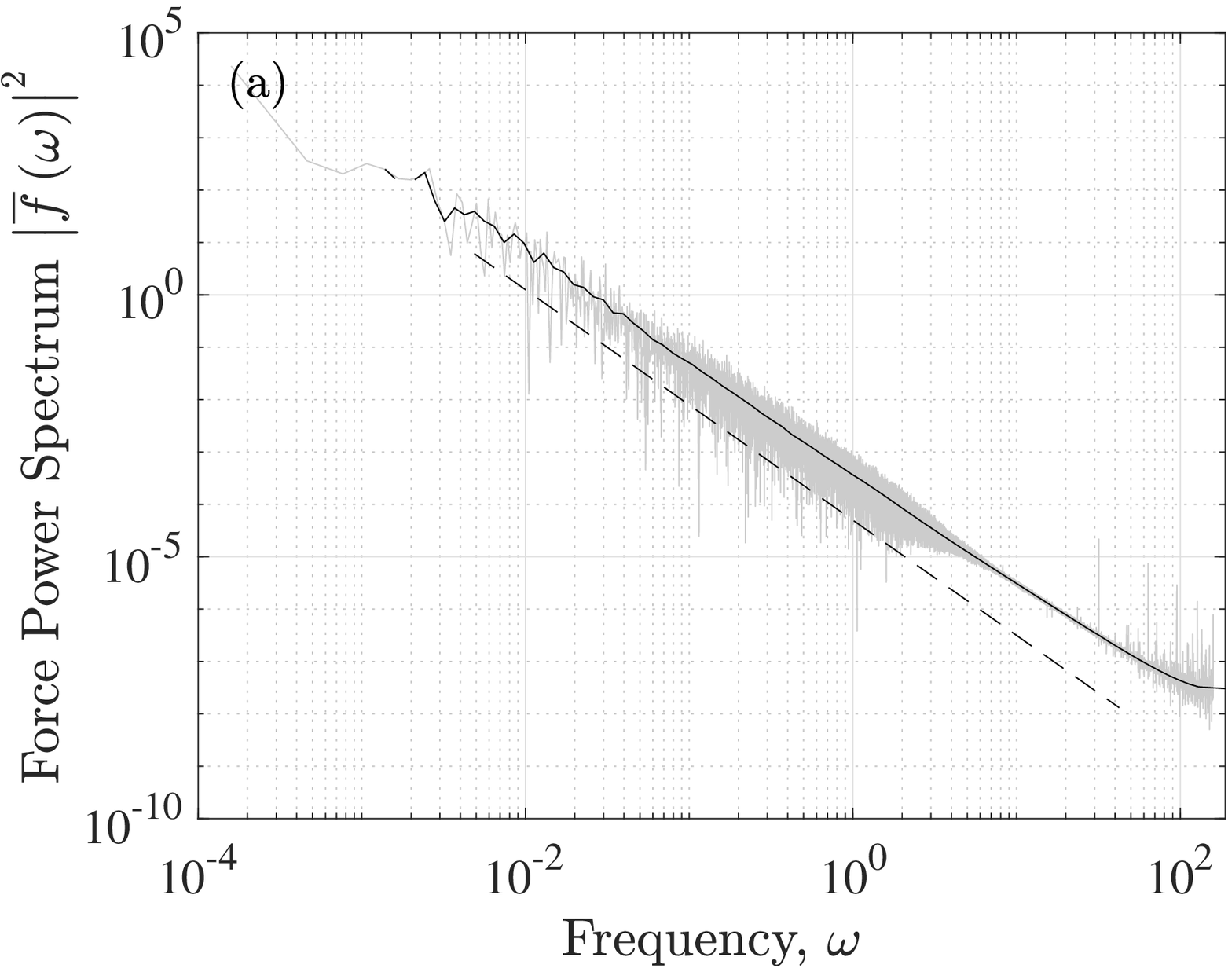}
         \centering
            \begin{tikzpicture}
              \coordinate (a) at (0,0); 
                \node[white] at (a) {\tiny.};               %
                \drawSlope{3}{2.9}{0.4}{54}{black}{\hspace{-2pt}2.2}{\vspace{2pt}1}
			\end{tikzpicture}

         \label{fig:PSDC0.1}
     \end{overpic}
     \begin{subfigure}[b]{\ratio\textwidth}
         \centering
         \includegraphics[width=\textwidth]{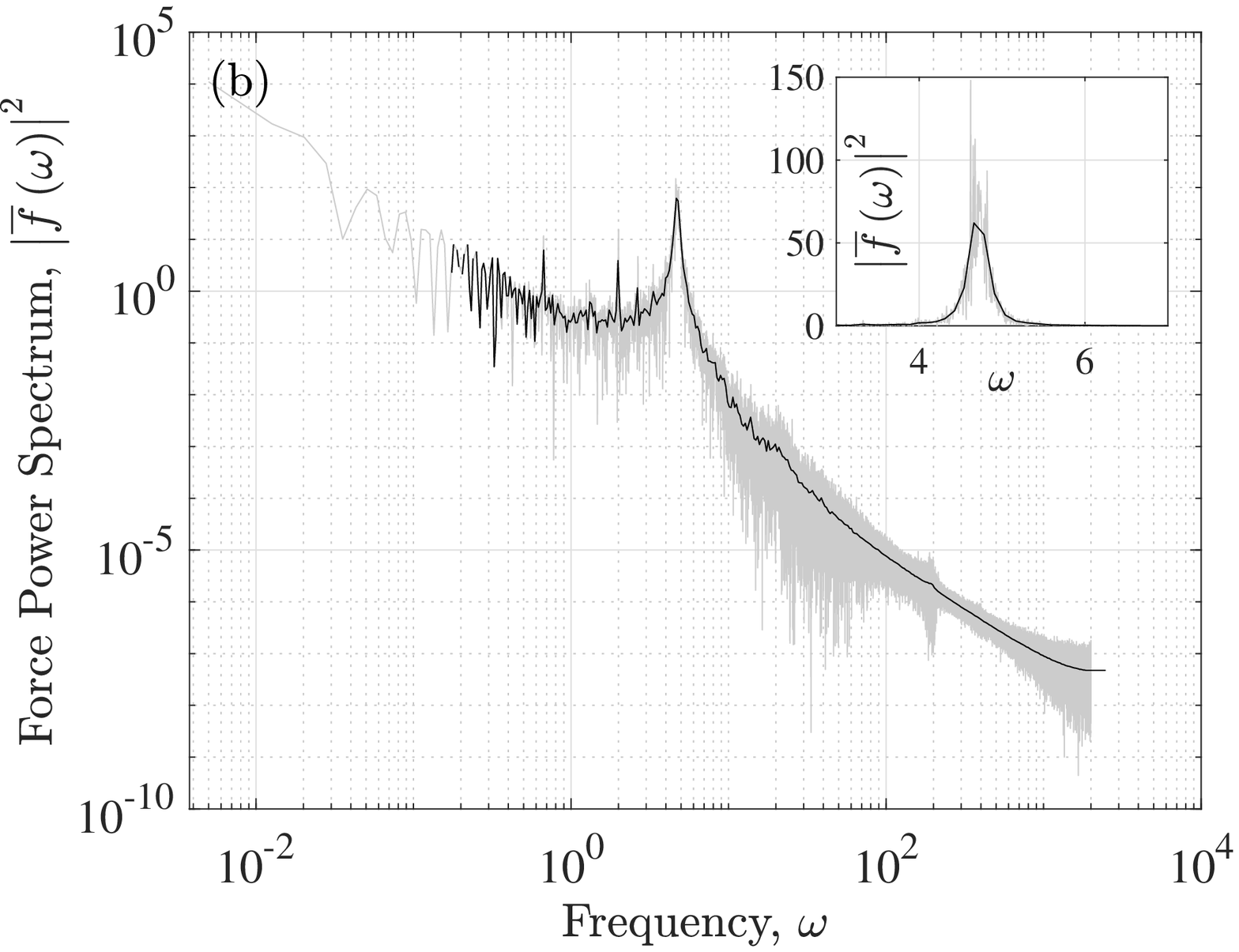}
         \label{fig:PSDC10}
     \end{subfigure}
        \caption{Power spectrum $|\bar f(\omega)|^2$ associated with the force timeseries at \textbf{a)} $c=0.1$ mm/s \textbf{b)} $c=1000$ mm/s. The inset for \textbf{b)} is the same as the main plot but on the lin-lin scale. The dashed line for \textbf{a)} denotes $|\bar f(\omega)|^2\propto \omega^{-\delta}$ with $\delta=2.2$. The solid curves correspond to the binned data.}
        \label{fig:powerSpectra}
\end{figure}

\section{\label{sec:Transition} Stick-slip dynamics vs. stable sliding}
We performed a series of tests on samples with different pulling speeds $c$. The magnitude of the resulting force $\mathbf{F}$ and velocity of the slider $\mathbf{v}$ scaled by $c$ are plotted against time in Fig.~\ref{fig:Transition}.
At the slowest rate $c=0.1$ mm/s, the response is characterized by abrupt force drops preceded by longer stress build-up periods as in Fig.~\ref{fig:Transition}(a).
Similarly, the slider velocity exhibits a stick-slip dynamics with quiescent periods that are frequently interrupted by short-lived active phases.
By contrast, Fig.~\ref{fig:Transition}(b), corresponding to $c=1000$ mm/s, shows a well-established quasi-periodic sliding regime. Both dynamical behaviors are in agreement with experimental observations~\cite{zadeh2019seismicity,zadeh2019crackling,cheng2101correlating}.

In frequency domain $\omega$, the stick-slip dynamics at the low driving rate is marked by a scale free power law behavior associated with the power spectral density of the force signal $|\bar f(\omega)|^2$ as displayed in Fig.~\ref{fig:powerSpectra}(a).
We find $|\bar f(\omega)|^2\propto \omega^{-\delta}$ with $\delta\simeq 2.2$ over at least four decades in $\omega$.
This is in close agreement with the measured exponent corresponding to the experimental setup with $\delta = 2.4 \pm 0.2$, where the power law extended over a shorter range of frequencies \cite{zadeh2019crackling}.
In Fig.~\ref{fig:powerSpectra}(b), the power spectrum associated with the faster driving rate at $c=1000$ mm/s develops a characteristic peak at $\omega\simeq 5~Hz$ which is a signature of the quasi-periodic signal in the time domain.
As reported in the experimental setting \cite{zadeh2019crackling}, this characteristic frequency should scale with the driving rate $c$ and differ from the natural frequency set by the slider mass $M_\text{slider}$ and (pulling) spring constant $k_\text{spring}$, i.e. $\omega^2=k_\text{spring}/M_\text{slider}$.
 It should be noted that the driving rate in Fig.~\ref{fig:powerSpectra}(b) is a factor of 10 larger than the fastest case reported in the experiment which leads to a more pronounced peak in the frequency domain. 

\section{\label{sec:Seimicity} ``Seimicity" analysis}
In this section, we investigate the properties of the stick-slip events of the slider, often denoted as avalanches. 
Specifically, we probe the dissipated energy power and its evolution with time to define the size and duration of individual avalanches. This allows us to quantify the avalanche size statistics along with inter-occurrence time distributions both showing non-trivial scaling features at low driving rates.
Other statistical measures such as productivity relation or temporal aftershock rates require a proper identification of mainshock-aftershocks sequences which will be discussed in subsequent sections.

\begin{figure}[t]
     \centering
      \begin{subfigure}[b]{\ratio\textwidth}
          \centering
          \includegraphics[width=\textwidth]{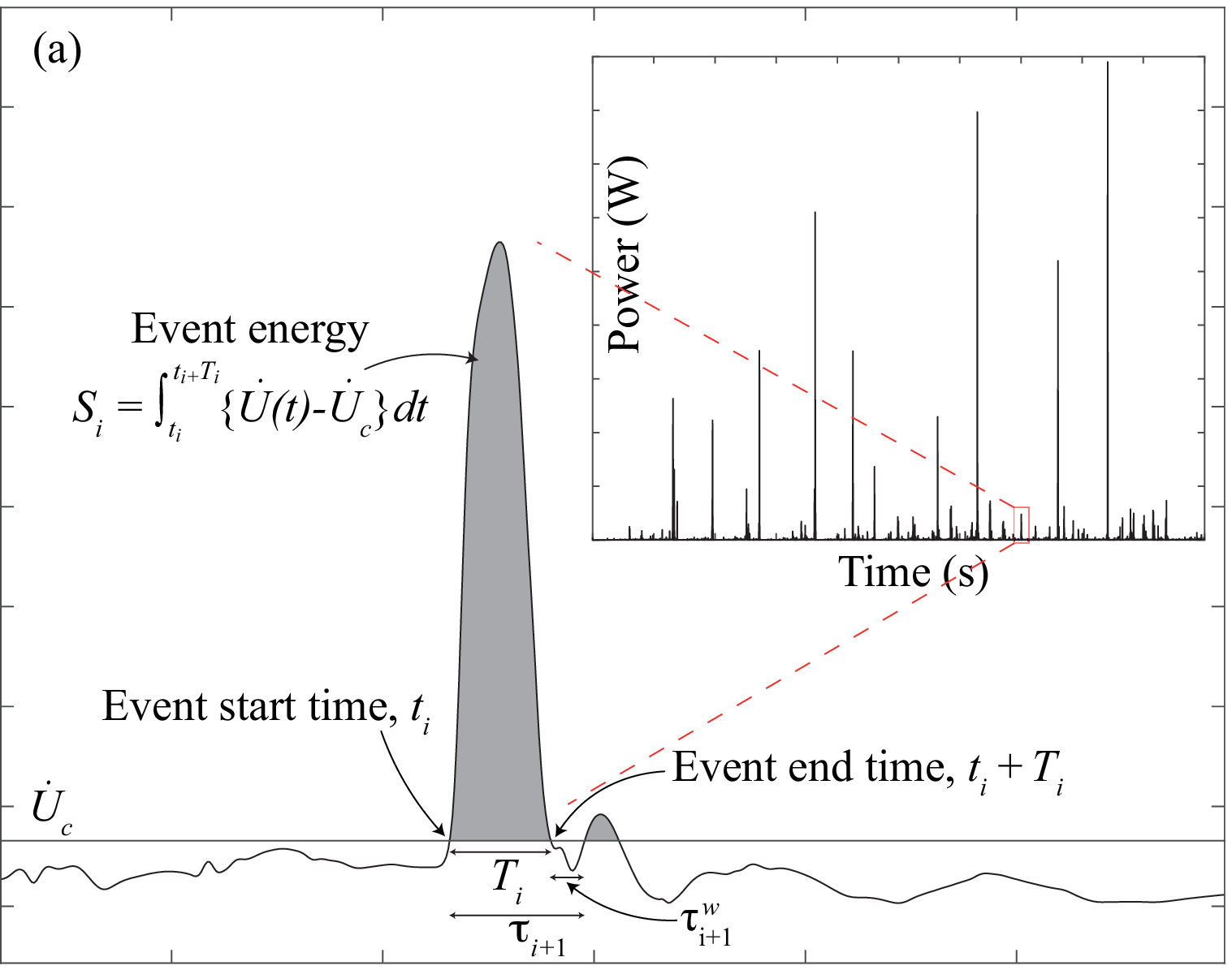}
          \includegraphics[width=\textwidth]{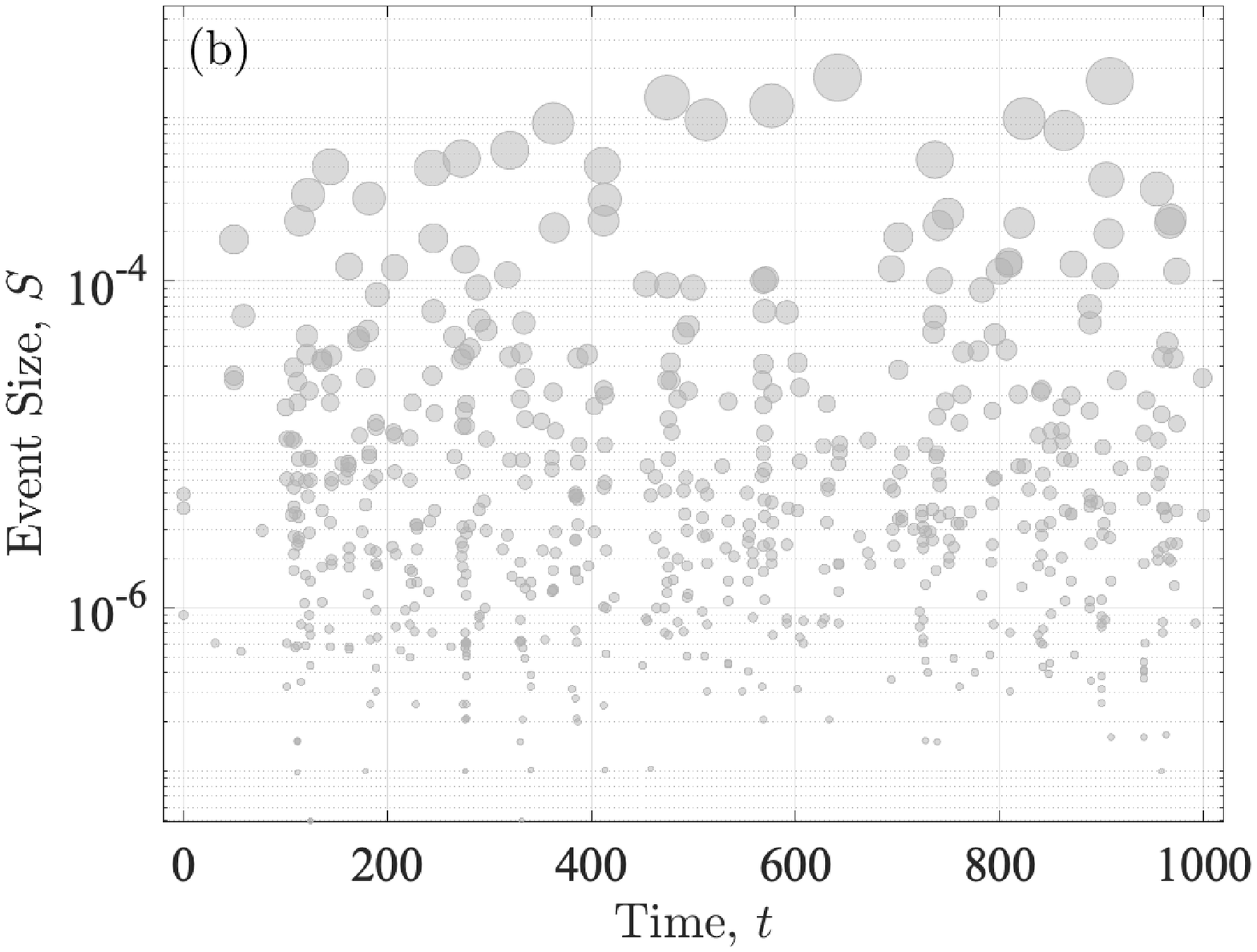}
      \end{subfigure}
         \caption{\textbf{a)} Definition of the avalanche size $S$, avalanche duration $T$, avalanche inter-occurrence time $\tau$, and slider rest time $\tau_w$ at $c=0.1~mm/s$. The main plot is a magnified view of the power time series shown in the inset. \textbf{b)} Magnitude time series.}
         \label{fig:EventDetails}
\end{figure}
\begin{figure}[b]
     \begin{overpic}[width=\ratio\textwidth]{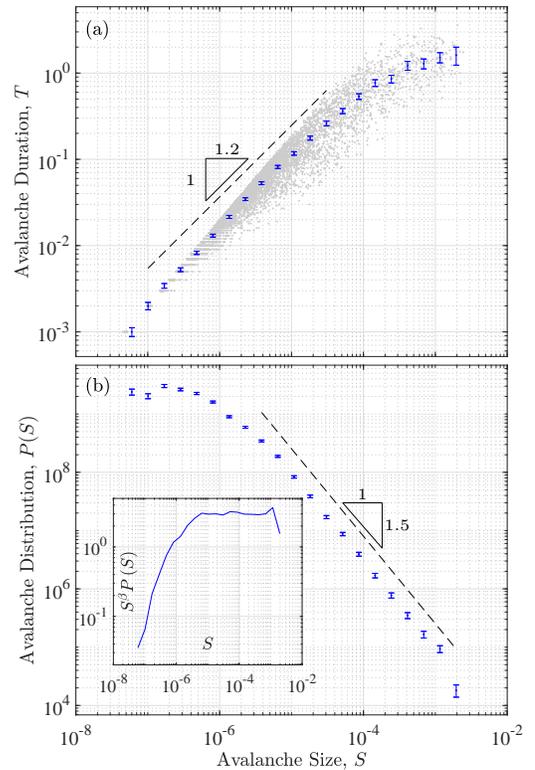}
         \centering
            \begin{tikzpicture}
              \coordinate (a) at (0,0); 
                \node[white] at (a) {\tiny.};  %
                \drawSlope{2.8}{8.0}{0.4}{-45}{black}{\hspace{0pt}1}{\vspace{-4pt}1.2}
                \drawSlope{4.6}{3.4}{0.4}{221}{black}{\hspace{0pt}1.5}{\vspace{-4pt}1}
			\end{tikzpicture}
     \end{overpic}
     \caption{\textbf{a)} Scatter plot of the avalanche duration $T$ vs. avalanche size $S$, \textbf{b)} avalanche size distributions, both at $c=0.1~mm/s$. The symbols in \textbf{a)} indicate the mean duration $\bar T$ over prescribed bins in $S$. The dashdotted line in \textbf{a)} indicates a power law $T\propto S^{1/\gamma}$ with $\gamma=1.2$. The dashdotted line in \textbf{b)} is a guide to the power law $P(S)\propto S^{-\beta}$ with $\beta = 1.5$. The error bars denote one standard error. The inset is the same as the main plot but rescaled by $S^{\beta}$.}
    \label{fig:EventSizeEventDuration}
\end{figure}

\subsection{\label{sec:Avalanche Statistics}Avalanche Statistics}
We define the spontaneous rate of dissipated energy of the slider as $\dot U=\mathbf{F}\cdot\mathbf{v}$ during the slip period.
The power signal is shown in Fig.~\ref{fig:EventDetails}(a) with a noise floor $\dot U_c \simeq 10^{-6}$ that is intermittently interrupted by short-lived stick-slip events or avalanches. 
Figure~\ref{fig:EventDetails}(b) displays the timeseries associated with the avalanche size $S=\int_{t_i}^{t_i+T_{i}} \{\dot{U}(t)-\dot{U_c}\}~dt$ which has dimensions of energy and corresponds to an avalanche initiated at $t_i$ with duration $T_{i}$. 

Figure~\ref{fig:EventSizeEventDuration} displays the scatter plot of the avalanche size $S$ and event duration $T$ along with the avalanche size distributions $P(S)$. 
The scatter plot in Fig.~\ref{fig:EventSizeEventDuration}(a) demonstrates that, statistically speaking, larger avalanches tend to have longer duration with a scaling behavior that may be described on average as $T\propto S^{1/\gamma}$ with $\gamma\simeq1.2$, in agreement with the experimental observation \cite{zadeh2019seismicity}.
The scaling regime spans almost three decades in $S$ before reaching a plateau at large avalanche sizes.
The avalanche size distribution in Fig.~\ref{fig:EventSizeEventDuration}(b) decays as a power-law $P(S)\propto S^{-\beta}$ over at least three decades above a (lower) cut-off size $S_\text{min}\simeq 10^{-6}$ with $\beta \simeq 1.5$ which is within the range of measured exponents in experiments $1.2-1.7$ \cite{lherminier2019continuously, zadeh2019seismicity} and matches the mean-field estimate $\beta=3/2$ \cite{fisher1998collective}.
In all of the subsequent avalanche analyses, we use $S_\text{min}=10^{-6}$ as a lower bound for event size thresholds, e.g. $S_c\ge S_\text{min}$, to ensure a meaningful (power-law) scaling regime associated with $S$.

The avalanche size distribution can be expressed in an accumulated form as $P(\text{size}>S)=(S_\text{min}/S)^{\beta-1}$. 
This distribution may also be transformed into the classical Gutenberg-Richter \emph{magnitude}-frequency relation $P(\text{mag.}>m)=10^{-b(m-m_c)}$ with magnitude of completeness $m_c$ and $b-$value that controls the exponential decay rate.
The magnitude $m$ is empirically related to seismic moment $M$ via $m= c_0\text{log}_{10}M-c_1$ with \emph{non}-universal parameters $c_0$ and $c_1$ \cite{kagan2013earthquakes}.
From this relationship, it follows that $b=(\beta_M-1)/c_0$ with exponent $\beta_M$ describing the power-law decay of seismic moments distributions \cite{de2016statistical}. 
It should be noted that we measure the dissipated energy $S$, not the seismic moment $M$ with the latter typically evaluated based on the shear modulus, slip size, and associated stress drop.
Provided that $S$ is proportional to $M$ (see \cite{de2016statistical, lord2020seismic} and references therein for a discussion of this assumption), one obtains $\beta=\beta_M$.
Using $c_0=\frac{2}{3}$ established for \emph{large} earthquakes~\cite{de2016statistical}, we obtain $b\simeq 0.75$ which is smaller than the commonly observed $b\approx 1$ in tectonic settings, though there is some variability across different settings~\cite{schorlemmer2005variations, gu2013triggering, scholz2015stress, davidsen2016self}. One could, however, dispute the conversion of $\beta$ to $b$-value for our small avalanche sizes here~\cite{ben2002potency,lord2020seismic}, such that a direct comparison with tectonic seismicity might not be appropriate.



\begin{figure}[t]
     \begin{overpic}[width=\ratio\textwidth]{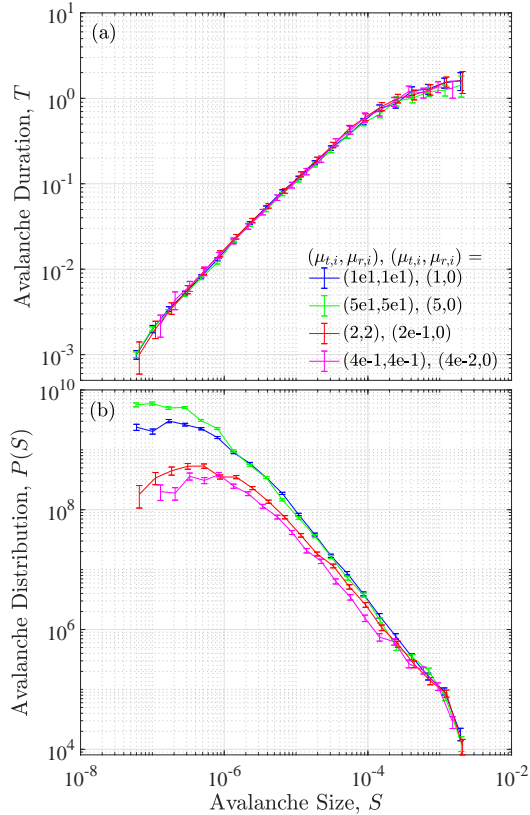}
         \centering
            \begin{tikzpicture}
              \coordinate (a) at (0,0); 
                \node[white] at (a) {\tiny.};  %
			\end{tikzpicture}
     \end{overpic}
         \caption{Avalanche statistics at multiple inter-particle sliding friction $\mu_{t}$ and rolling friction $\mu_{r}$. \textbf{a)} Avalanche duration $T$ plotted against event size $S$. \textbf{b)} (non-normalized) avalanche size distribution $P(S)$. The first pair of parameters correspond to inter-particle friction and second one relates to the slider-substrate friction.}
         \label{fig:avlFriction}
\end{figure}

\begin{figure}[t]
     \begin{overpic}[width=\ratio\textwidth]{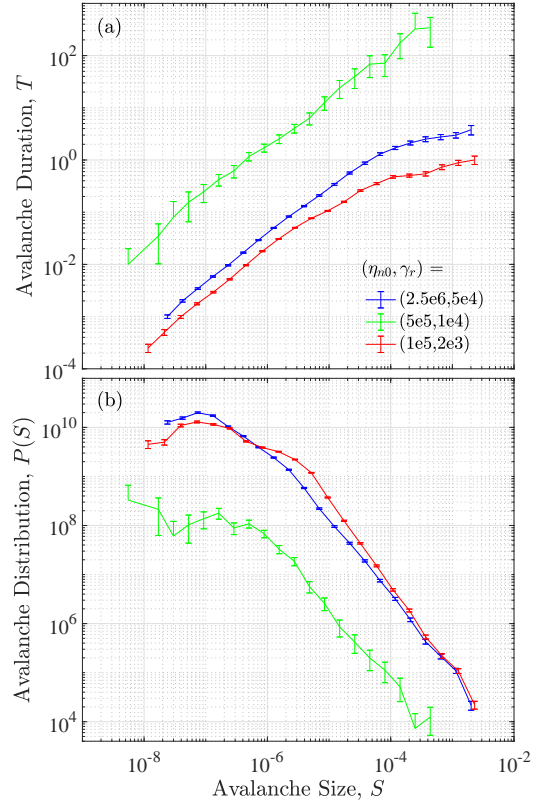}
         \centering
            \begin{tikzpicture}
              \coordinate (a) at (0,0); 
                \node[white] at (a) {\tiny.};  %
			\end{tikzpicture}
     \end{overpic}
         \caption{Avalanche statistics at multiple damping rates for both normal and rolling components $\eta_{n0}, \gamma_{r}$. \textbf{a)} Avalanche duration $T$ versus event size $S$ \textbf{b)} (non-normalized) avalanche size distribution $P(S)$.}
         \label{fig:avlDamping}
\end{figure}

\subsubsection{Robustness Analysis}

Figure~\ref{fig:avlFriction} and \ref{fig:avlDamping} examine the overall robustness of our findings with respect to changes in the inter-particle (sliding and rolling) friction coefficient as well as the (normal and rolling) dissipation time scales.
Displayed in Fig.~\ref{fig:avlFriction}(a), the scaling relation between avalanche size and duration remains almost insensitive to variations in the microscopic friction.
The avalanche size exponent $\beta$ seems to be also a robust scaling feature at larger $S$ values as shown in Fig.~\ref{fig:avlFriction}(b).
Higher damping rates in Fig.~\ref{fig:avlDamping}(a) lead to an overall shift in the avalanche duration but has no discernible effect on exponent $\gamma$ within the scaling regime at small and intermediate $S$.
This is also true for size distributions in Fig.~\ref{fig:avlDamping}(b) where $\beta\simeq 1.5$ seems to be a robust scaling exponent.

In order to probe potential effects of the surface roughness of the slider on the avalanche statistics, we considered different cases of depths of the grooves at the base of the slider on the granular substrate.
We find that there are no considerable effect on the overall avalanche statistics (see Fig.~\ref{fig:Slider} ). In particular, the power-law exponents are unaltered indicating that the variation in the interaction between the slider and the substrate do not affect the universality class. 
Similarly, changes in polydispersity (see Fig.~\ref{fig:Polydis}) do not lead to any quantitative changes between the different sets of avalanche size and duration distributions.

\subsection{\label{sec:OccurrenceTimes} Avalanche Inter-occurrence Time and Slider Rest Time Statistics }
\begin{figure}[t]
     \centering
     \begin{overpic}[width=\ratio\textwidth]{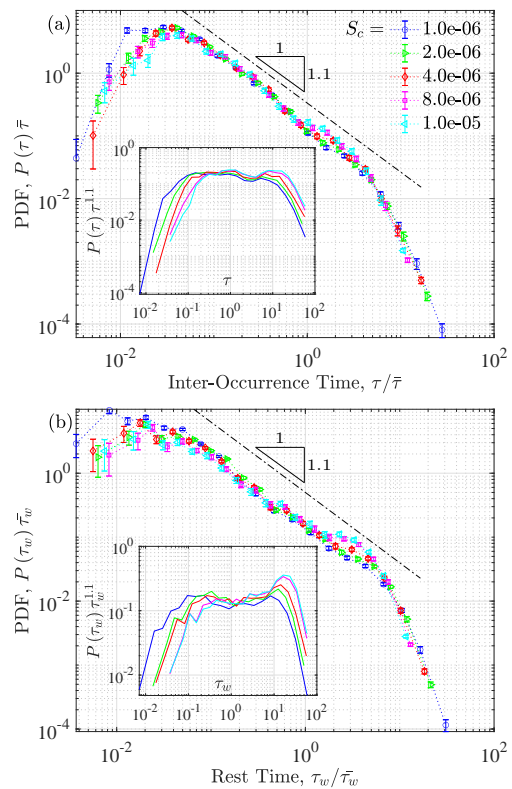}
         \centering
            \begin{tikzpicture}
              \coordinate (a) at (0,0); 
                \node[white] at (a) {\tiny.};
                \drawSlope{3.5}{9.85}{0.4}{234}{black}{\hspace{0pt}1.1}{\vspace{-4pt}1}
                \drawSlope{3.5}{4.65}{0.4}{234}{black}{\hspace{0pt}1.1}{\vspace{-4pt}1}
			\end{tikzpicture}
	\end{overpic}
    \caption{Distribution of avalanche inter-occurrence time, $\tau$, and slider rest time, $\tau_{w}$ at multiple threshold sizes, $S_c$ at $c=0.1$ mm/s. \textbf{a)} Event inter-occurrence time distribution, $P(\tau)\bar{\tau}$ versus $\tau/\bar{\tau}$ with ${\bar\tau}$ denoting the mean inter-occurrence time with $S > S_c$. \textbf{b)} Slider rest time distribution, $P(\tau_w)\bar{\tau}_w$ versus $\tau_{w}/\bar{\tau_{w}}$, similar to the previous case, ${\bar\tau}_w$ denotes the mean rest time. The dash-dotted line on both the plots indicates a power-law decay $P\{\tau_{(w)}\}\propto \tau_{(w)}^{-1.1}$. The insets show $P\{\tau_{(w)}\}\times\tau_{(w)}^{1.1}$. The error bars denote one standard error.} 
    \label{fig:WaitingTime} 
\end{figure}

For a homogeneous Poisson process with independent events of constant rate $\lambda$, the inter-occurrence time statistics should obey an exponential distribution $P(\tau)=\lambda e^{-\lambda\tau}$ with $\tau_i=t_i-t_{i-1}$ denoting the time interval between two consecutive avalanches, i.e., the time interval between the onset of avalanche $i-1$ and the onset of avalanche $i$ (see Fig.~\ref{fig:EventDetails}(a)).
Deviations from this exponential distribution are demonstrated in Fig.~\ref{fig:WaitingTime}(a) where the rescaled distributions $\bar\tau P(\tau)$ are plotted against $\tau/\bar \tau$ for avalanches of sizes $S > S_c$.
Here $\bar\tau^{-1}$ denotes the occurrence frequency of events with sizes larger than $S_c$. 
The rescaled distributions in the main plot are characterized by a power-law decay extending for almost two decades up to a slight hump at $\tau>\bar\tau$ that tends to become more pronounced with increasing $S_c$.
As shown in the inset of Fig.~\ref{fig:WaitingTime}(a), we find $P(\tau)\propto \tau^{-1.1}$ for the power-law decay which is consistent with the observed behavior in the spring-slider experiment with $P(\tau)\propto \tau^{-(1.1\pm 0.08)}$ \cite{zadeh2019seismicity}. 

We also probed the statistics of slider rest times $\tau_{i}^{w} = \tau_i-T_{i-1}$ indicating the \emph{rest} period of the slider between the end time of avalanche ${i-1}$ and start time of avalanche $i$ (see Fig.~\ref{fig:EventDetails}(a)).
Here $T_{i-1}$ denotes the avalanche duration associated with event $i-1$.
In the context of a dynamical point process, a typical assumption is that $T_{i-1}\ll \tau_i$ and, therefore, differences between $P(\tau)$ and $P(\tau_w)$ shall be statistically insignificant. In the present set-up, however, the time-scale separation is not directly applicable since large avalanches have a duration significantly longer than the shortest observed inter-occurrence time as a comparison of Fig.~\ref{fig:EventSizeEventDuration}(a) and Fig.~\ref{fig:WaitingTime}(a) shows.
Nevertheless, Fig.~\ref{fig:WaitingTime}(b) and the inset illustrate that the rest time distributions decay almost identical to the inter-occurrence time distribution, e.g. $P(\tau_w)\propto \tau_w^{-1.1}$.


\subsection{\label{sec:TemporalCorrelations}  Correlation Analysis}
\begin{figure}[t]
     \begin{overpic}[width=0.5\textwidth]{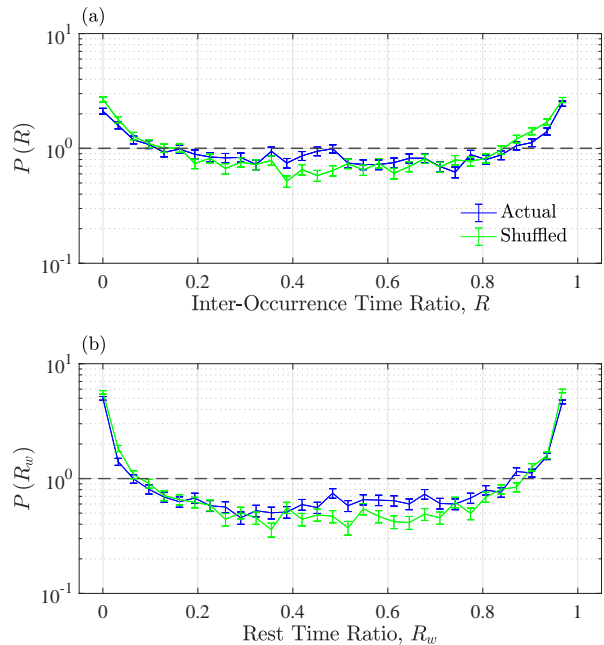}
         \centering
     \end{overpic}
         \caption{Statistical distribution \textbf{a)} $P(R)$ of the inter-occurrence time ratio $R$ for actual and shuffled catalogs, \textbf{b)} $P(R_w)$ of the rest time ratio $R_w$ for actual and shuffled catalogs, both at threshold size $S_{c}=1.0\text{e}-6$, and $c=0.1$ m/s. The flat dashed lines denote a (potentially non-homogeneous) Poisson process. The error bars denote one standard error.}
         \label{fig:Rthresholds}
\end{figure}

To understand the similarities between the distributions of inter-occurrence times and rest times, a correlation analysis is helpful. First, we analyze temporal auto-correlations of these quantities separately by probing fluctuations in $R_{(w)}\doteq{\tau^{(w)}_{i+1}}/(\tau^{(w)}_{i+1}+\tau^{(w)}_i)$ defined as the ratio between successive time lags within the respective sequence \cite{van2010connecting}. For both homogeneous and non-homogeneous Poisson processes, $P\{R_{(w)}\} \equiv 1$ for $0<R_{(w)}<1$ such that a uniform distribution would indicate an absence of correlations. Instead, the two distributions of $R$ and $R_{w}$ shown in Fig.~\ref{fig:Rthresholds}(a), (b) exhibit strong deviations from a uniform distribution with two distinct peaks at $R_{(w)}\simeq 0$ and $1$.
Thus, we can also rule out that the temporal behavior is following a non-homogeneous Poisson process --- we had already established that both inter-occurrence times and rest times do not obey a homogeneous Poisson process since they exhibit broad distributions (see Fig.~\ref{fig:WaitingTime}). Yet, it is possible that the behavior of $P\{R_{(w)}\}$ is solely determined by these broad distributions. If indeed true, the observed behavior would not be indicative of correlations but rather of a renewal process. To test this alternative hypothesis, Fig.~\ref{fig:Rthresholds} also shows $P\{R_{(w)}\}$ for the shuffled magnitude timeseries, where the order of the inter-occurrence times is randomized such that all auto-correlations are destroyed. The same is true for the associated rest times --- note that each inter-occurrence time is the sum of the event duration $T$ and the rest time $\tau_w$. In both cases, Fig.~\ref{fig:Rthresholds} shows that there are no pronounced differences between the original case and the shuffled case. Thus, the abundance of low and high $R_{(w)}$ values is not indicative of correlations in this case but consistent with a renewal process.

\begin{figure}[t]
     \begin{overpic}[width=\ratio\textwidth]{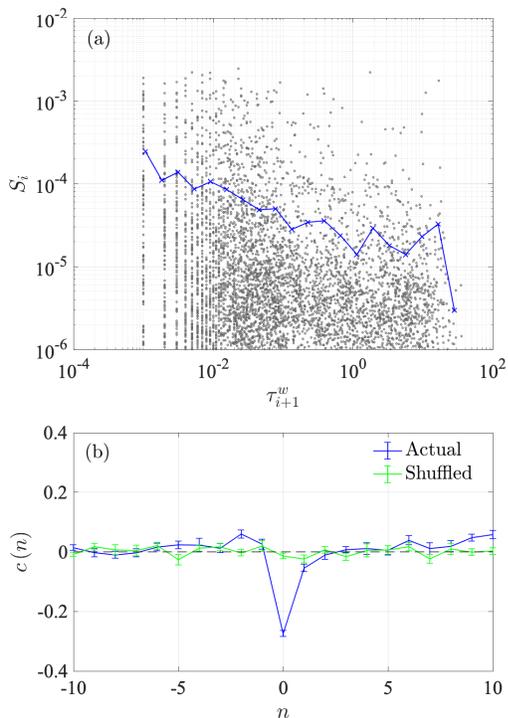}
         \centering
     \end{overpic}
     \caption{Cross-correlation analysis between avalanche sizes $S_i$ and rest times $\tau^w_{i+1}$. \textbf{a)} Scatter plot of $S_i$ and $\tau^w_{i+1}$ with $i$ being the avalanche index. \textbf{b)} Cross correlation function $c(n)$ with index shift $n$ evaluated for the actual and randomized catalogs. The (blue) curve in a) indicates a binning average. The (black) base line in b) indicates zero correlations. The error bars indicate one standard error.}
         \label{fig:crltnFunction}
\end{figure}

We also carried out a cross-correlation analysis between avalanche sizes $S_i$ and rest times $\tau^w_{i+1}$ as displayed in the scatter plot of Fig.~\ref{fig:crltnFunction}(a).
The plot exhibits a large scatter in the data but the observed trend indicates discernable anti-correlations between the logarithm of the two observables $x=\text{log}_{10}{S}$ and $y=\text{log}_{10}\tau_w$ with the correlation coefficient $\langle~\hat{x}_{i}.\hat{y}_{i+1}~\rangle_i\simeq -0.3$. 
Here $\langle.\rangle_i$ denotes averaging over the avalanche index $i$ and $\hat{x}$ indicates the fluctuating part (with the mean value subtracted) normalized by the standard deviation associated with each variable.
A negative correlation implies that high-energy events --- typically corresponding to events of long duration (see Fig.~\ref{fig:EventSizeEventDuration}(a)) --- tend to be followed by short rest times potentially indicative of main shock-aftershock dynamics \cite{makinen2015avalanches}. These negative correlations might also explain why the distributions of inter-occurrence times and rest times in Fig.~\ref{fig:WaitingTime} are very similar.
It is noteworthy that these negative correlations are not present in the shuffled catalog (data not shown).

We repeated the above analysis for multiple index shifts $n$ represented by the cross correlation function $c(\pm n)=\langle~\hat{x}_{i}.\hat{y}_{(i+1)\pm n}~\rangle_i$ with $n\in(0,1,2,...)$.  
As shown in Fig.~\ref{fig:crltnFunction}(b), $c(n)$ corresponding to the actual catalogs is basically indistinguishable from the noise floor (as indicated by the shuffled sequences) for $n\neq 0$.  
This suggests very localized ``memory" indicating that if main shock-aftershock dynamics is indeed present, larger events might only induce very few aftershocks, which is opposed to tectonic seismicity, where strong clustering is a hallmark of aftershock activity~\cite{gu2013triggering,davidsen2016self,davidsen2015generalized, zaliapin2013earthquake}. 
The observed anti-correlation for $n=0$ in the simulation seems to be a robust feature of the stick-slip dynamics showing insignificant variations with the slider roughness (see Fig.~\ref{fig:crRoughness}).
With increasing smoothness, however, a positive peak appears at $n=-1$ implying that there is a tendency that long (short) resting times are followed by large (small) avalanche sizes (see Fig.~\ref{fig:crRoughness}).
This might be indicative of the transition to the quasi-periodic (stable) sliding behavior but within the quasi-static regime.
These positive correlations tend to persist at finite driving rates on rough sliders (data not shown) suggesting that both rate effects and roughness features control this transitional behavior.




\subsection{\label{sec:AfterhocksAnalysis}Aftershocks Analysis}
To directly analyze if main shock-aftershock dynamics is present, we follow a simple methodology to identify aftershocks \cite{baro2013statistical}, which also has been used to analyze the related lab experiments~\cite{zadeh2019seismicity,zadeh2019crackling}. 
In this framework, an aftershock sequence triggered by the $i$-th avalanche (or \emph{main shock}) of magnitude $S^{(i)}_\text{ms}$ with $N$ subsequent aftershocks is defined as
\begin{equation}\label{eq:sequence}
\{S^i_\text{ms};~S_{i+1}<S^i_\text{ms},~...,~S_{i+N}<S^i_\text{ms},~S_{i+N+1}>S^i_\text{ms}\}.
\end{equation}
Furthermore, we repeat the aftershock analysis for the \emph{shuffled} timeseries --- generated by a random permutation of the order of avalanche sizes $S$, event duration $T$, and rest times $\tau_w$. The latter two quantities are shuffled in phase with each other in order to maintain the actual inter-occurrence times $\tau$ since $\tau_i = T_{i-1}+\tau^{i}_w$ with $i$ being the avalanche index. This allows us to evaluate the role of inter-event correlations and serves as a simple \emph{null} model of trivial ``aftershocks". 

\subsubsection{\label{sec:productivity}Aftershock productivity relation}
\begin{figure}[t]
     \centering
     \begin{overpic}[width=\ratio\textwidth]{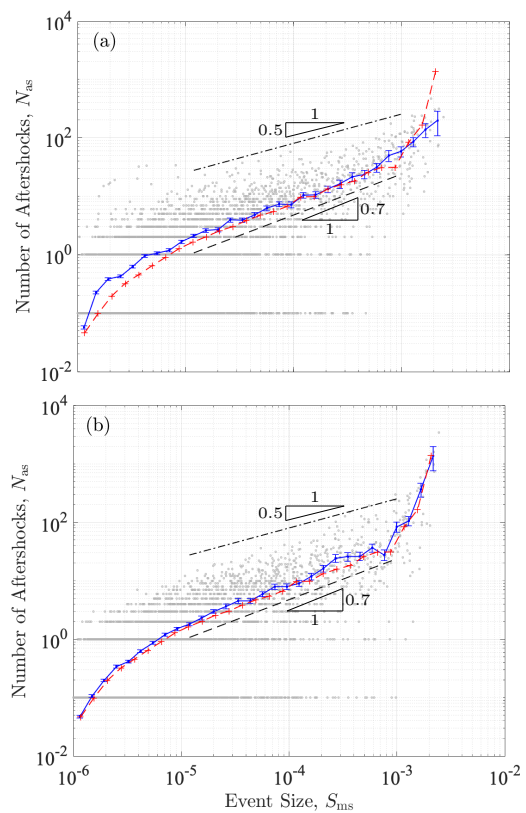}
         \centering
            \begin{tikzpicture}
              \coordinate (a) at (0,0); 
                \node[white] at (a) {\tiny.};
                \drawSlope{4.0}{9.2}{0.4}{-76}{black}{\hspace{0pt}0.5}{\vspace{-4pt}1}
                \drawSlope{4.2}{8.15}{0.4}{112}{black}{\hspace{0pt}0.7}{\vspace{-4pt}1}
                
                \drawSlope{4.0}{4.1}{0.4}{-76}{black}{\hspace{0pt}0.5}{\vspace{-4pt}1}
                \drawSlope{4.0}{2.95}{0.4}{112}{black}{\hspace{0pt}0.7}{\vspace{-4pt}1}
			\end{tikzpicture}
	\end{overpic}
         \caption{Number of aftershocks $N_\text{as}$ plotted against the trigger size $S_\text{ms}$ corresponding to the \textbf{a}) actual \textbf{b}) shuffled time series with $S_c=10^{-6}$. The blue curves indicate the mean value $\bar{N}_\text{as}$ over prescribed magnitude bins. The straight lines indicate the productivity relation $\bar{N}_\text{as} \propto S_\text{ms}^{\alpha}$ with $\alpha = 0.7$ (dashed lines below the curves) and $\alpha = \beta - 1$ (dashed-dotted lines above the curves). Here $\beta$ denotes the avalanche size exponent. The data points at the base of both panels indicate those events that did not trigger any aftershocks. The red curves on both panels indicate the theoretical prediction based on the accumulative event size distribution (see the main text).}
         \label{fig:Productivity}
\end{figure}

We first turn to the variation in the number of triggered aftershock events $N_\text{as}$ associated with the triggers of size $S_\text{ms}$ as displayed in Fig.~\ref{fig:Productivity}.
The scatter plot also includes the mainshocks that have no aftershocks.
In Fig.~\ref{fig:Productivity}(a), the average number of aftershocks $\bar{N}_\text{as}$ features a power-law scaling with size, \emph{i.e.} $\bar N_\text{as}\propto S_\text{ms}^{\alpha}$ with $\alpha \simeq 0.7$, shown as the (black) dashed line style, known as the productivity exponent in the context of the tectonic seismicity \cite{wetzler2016regional}.

Figure.~\ref{fig:Productivity}(b) compares the productivity data associated with the shuffled sequences leading to similar scaling features as observed in Fig.~\ref{fig:Productivity}(a). This strongly suggests that the observed behavior including the value of $\alpha$ can be derived under the assumption of \emph{independent} events.
In fact, based on the used methodology to identify aftershocks and the assumption of independent events, the productivity relation may be uniquely determined by the accumulated distribution of the avalanche size as reported in \cite{bares2018aftershock}.
In this context, $\bar{N}_\text{as}=F(S)/[1-F(S)]$ where $F(S)=P(\text{size} < S)$.
The theoretical relation is shown in Fig.~\ref{fig:Productivity}(a) (dashed red curve) which closely predicts the mean aftershock number. Furthermore, assuming a pure power law form for $P(S)$ with $F(S)=1-(S_\text{min}/S)^{\beta-1}$, it follows that $\bar N_\text{as}\propto S_\text{ms}^{\beta-1}$ or $\alpha=\beta-1$ as illustrated by the (black) dashed-dotted line style in Fig.~\ref{fig:Productivity}(a). The deviation of this scaling relation $\alpha \approx 0.5$ from the observed behavior $\alpha \approx 0.7$ is due to the fact that $P(S)$ (see the inset of Fig.~\ref{fig:EventSizeEventDuration}(b)) includes all small avalanches that do not follow a power law behavior.

It should be noted that both the shuffled and the actual data sets have the exact same accumulative size distribution $F(S)$, which leads to the identical prediction for the magnitude scaling of $\bar{N}_\text{as}$. The above observations validate the independence assumption which is based upon the absence of notable (magnitude) correlations between events. Moreover, they indicate that the ``productivity" relation is fully determined by the avalanche size distribution $P(S)$, which is in sharp contrast to (tectonic) seismicity~\cite{bares2018aftershock,daschercontrols,wetzler2016regional,marsan2017variable}. 


\begin{figure}[b]
     \centering
     \begin{overpic}[width=\ratio\textwidth]{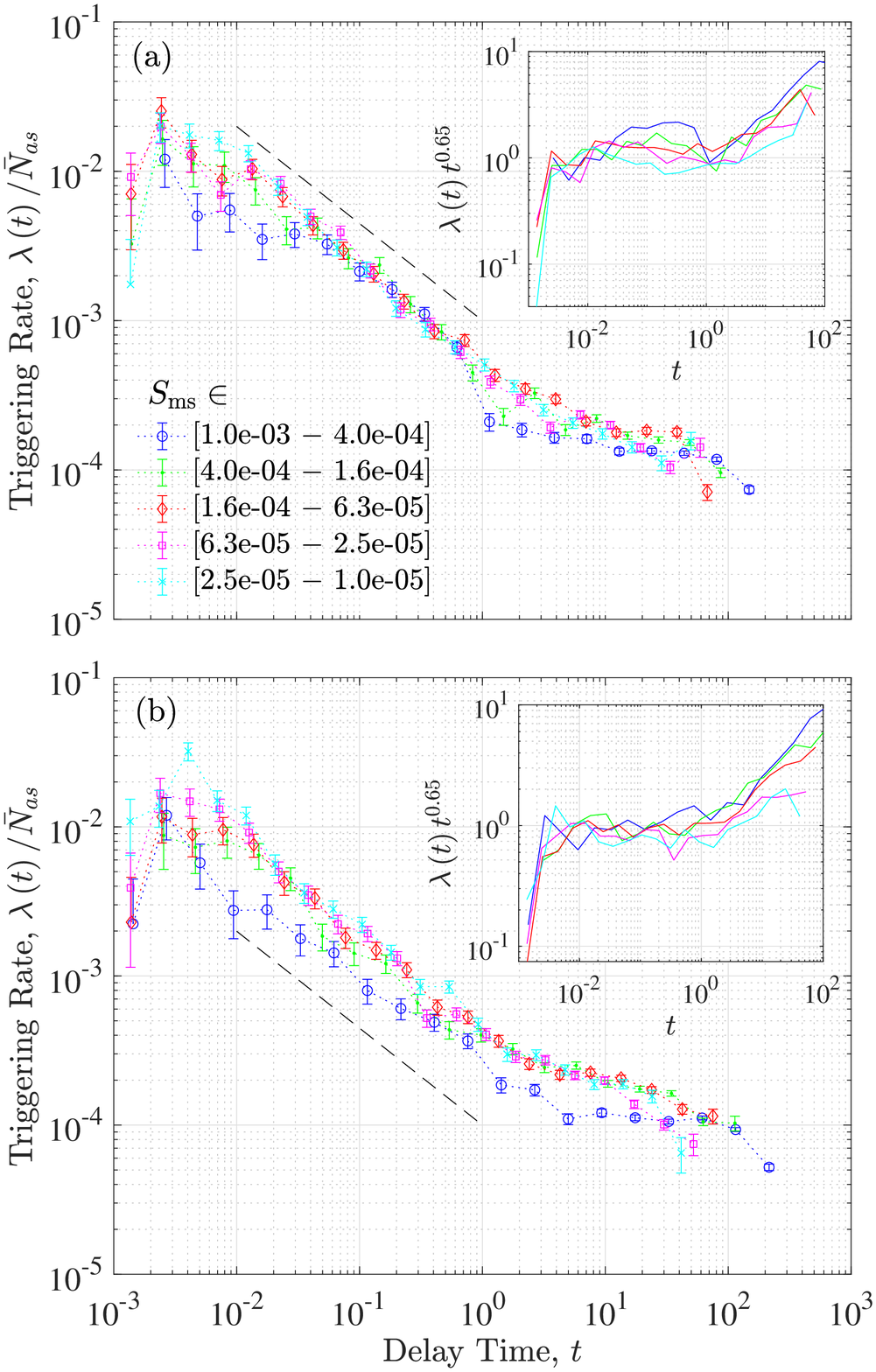}
         \centering
            \begin{tikzpicture}
              \coordinate (a) at (0,0); 
                \node[white] at (a) {\tiny.};
                \drawSlope{2.1}{9.6}{0.4}{231}{black}{\hspace{5pt}0.65}{\vspace{-4pt}1}
                \drawSlope{2.7}{2.7}{0.4}{52}{black}{\hspace{-5pt}0.65}{\vspace{-4pt}1}
			\end{tikzpicture}
	\end{overpic}
    \caption{Conditional triggering rate $\lambda(t)$ normalized by the mean number of aftershocks $\bar{N}_{as}$, versus delay time $t$ for multiple trigger sizes $S_\text{ms}$ corresponding to the \textbf{a}) actual \textbf{b}) shuffled time series. The (black) dashed line indicates $t^{-p}$ with $p=0.65$. The insets are the same as the main plots but rescaled by $t^{-p}$.}
         \label{fig:Omori}
\end{figure}

\subsubsection{\label{sec:aftershockRate} Temporal aftershock rates}
In the context of tectonic seismicity, 
an earthquake typically leads to an immediate increase in local seismic activity. The activity then decays algebraically with the delay time $t$ in accordance with the Omori-Utsu aftershock rate $\lambda(t)\propto t^{-p}$ \cite{utsu95} with scaling exponent $p$ typically estimated to be around unity \cite{scholz2002mechanics}.
The aftershock rates conditioned on the trigger size $S_\text{ms}$ and rescaled by $\bar{N}_\text{as}$ are plotted in Fig.~\ref{fig:Omori}(a).
It should be noted that $\int_t \lambda(t)~dt = \bar{N}_\text{as}$ such that the rescaling of the aftershock rates amounts to a separation of the productivity relation from the Omori-Utsu relation. As the data collapse shows, the rescaled aftershock rates are largely independent of the trigger size.
%
We find $p \simeq 0.65$ which seems to capture a robust scaling regime over at least two decades in time. 
Figure~\ref{fig:Omori}(b) shows the aftershock triggering rate of the shuffled data. 
They are largely indistinguishable from the unshuffled actual data in Figure~\ref{fig:Omori}(a), and the decrease in the rate can still be scaled as $\lambda(t)\propto t^{-p}$, with $p \simeq 0.65$ holding true as a good scaling exponent.
This implies that the temporal "aftershock" rates including the value of $p$ are fully determined by the inter-occurrence time statistics, which is the same for both the shuffled and the actual catalog. This is again in sharp contrast to tectonic seismicity, where aftershocks are a reflection of strong space-time-magnitude correlations~\cite{davidsen2015generalized,davidsen2016self,shcherbakov2005model}. 

\begin{figure}[t]
     \centering
     \begin{overpic}[width=\ratio\textwidth]{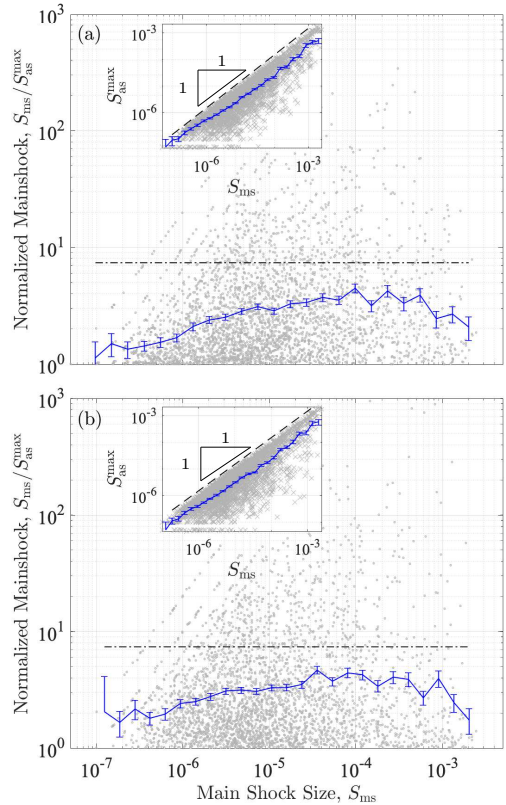}
         \centering
            \begin{tikzpicture}
              \coordinate (a) at (0,0); 
                \node[white] at (a) {\tiny.};%
                \drawSlope{2.8}{9.65}{0.4}{-53}{black}{\hspace{0pt}1}{\vspace{-4pt}1}
                \drawSlope{2.85}{4.65}{0.4}{-56}{black}{\hspace{0pt}1}{\vspace{-4pt}1}
			\end{tikzpicture}
	\end{overpic}
         \caption{Scatter plot of the normalized mainshock energy $S_\text{ms}/S_\text{as}^\text{max}$ and mainshock size $S_\text{ms}$ corresponding to \textbf{a}) actual \textbf{b}) randomized sequences. The insets plot the maximum aftershock size $S_\text{as}^\text{max}$ within a mainshock-aftershock sequence against the corresponding mainshock energy $S_\text{ms}$. The curves denote the bin-averaged data. The dashed lines in the insets indicate the B\r{a}th's relation $S_\text{as}^\text{max}\propto S_\text{ms}$.}
         \label{fig:Bath}
\end{figure}

\subsubsection{\label{BathLaw} B\r{a}th's relation}
B\r{a}th's relation describes the relative magnitude of aftershocks based on empirical observations that the magnitude difference between a mainshock and its largest aftershock is approximately constant, independent of the main shock magnitude \cite{baath1965lateral}.
Using the energy scale, this implies a constant ratio between the main shock size $S_\text{ms}$ and the maximum energy within the associated aftershock sequence  $S_\text{as}^\text{max}$.
The scatter plot corresponding to these two quantities is given in Fig.~\ref{fig:Bath}(a) and shows that B\r{a}th's relation holds quite well. 
Shuffling the magnitude time series does not seem to result in any qualitative change in B\r{a}th's relation (see Fig.~\ref{fig:Bath}(b)), similar to the other main relations related to aftershocks discussed in the preceding sections. This indicates that the origin of B\r{a}th's relation here can be understood based on independent events.
In this framework, one can derive that $S_\text{ms}/ S_\text{as}^\text{max}=e^{1/(\beta-1)}$ (see the appendix for further details). Yet, this simple argument tends to overestimate the actual (numerical) ratio between $S_\text{ms}$ and $S^\text{max}_\text{as}$ by a factor of almost two.
It is likely that the presence of the lower and upper cut-offs in the avalanche size distributions (Fig.~\ref{fig:EventSizeEventDuration}(b)) may lead to this discrepancy similar to Fig.~\ref{fig:Productivity}, where the theoretical exponent slightly underestimates the actual productivity relationship which is, otherwise, predicted accurately based on the full cumulative distribution of avalanche sizes.

\section{Discussions}\label{sec:discussions}
Our numerical setup has closely replicated the empirical observations on the relevance of stick-slip dynamics and principal tectonic seismicity relations in the context of a spring-slider experiment carried out on a granular substrate \cite{zadeh2019crackling,zadeh2019seismicity}.
The observed critical features (under quasi-static loading) may be closely described by the Gutenberg-Richter frequency-magnitude distribution and the aftershock productivity relation as well as the Omori-Utsu aftershock rate and B\r{a}th's relation.
We find that the associated scaling exponents closely match the experimental estimates in \cite{zadeh2019crackling,zadeh2019seismicity} with the avalanche size exponent that is fairly consistent with mean-field predictions ($\beta=\frac{3}{2}$) interpreting the avalanche or stick-slip dynamics of the slider as a critical branching process \cite{fisher1998collective}. 
Our analysis also indicates the robustness of the estimated exponents over a wide range of relevant model parameters including the grain-level damping ratio, (sliding and/or rolling) friction, surface roughness, and polydispersity. 
Instead, another numerical shearing study (with a significantly different setup) found that the degree of polydispersity controls the decay of the distribution of avalanche size --- as measured by the stress drop --- with the strongest degree of polydispersity leading to a mean-field exponent~\cite{ma2020size}. It is likely that surface effects and roughness features (as opposed to polydispersity as a bulk property) dominate stick-slip dynamics within our spring-slider experiment. Indeed, both surface effects and roughness features are missing or are secondary features in the aforementioned simulations~\cite{ma2020size}. This is in line with observations in \cite{baro2021quasistatic} where avalanche sizes, depending on their notion as a surface or bulk property, were shown to follow different sets of statistics in a driven solid. 
We note that none of the above features are easily tunable in real experimental settings which in turn restricts investigation of the universality and generality of empirical findings and/or identification of essential control parameters.

Regarding the underlying origin of the aftershock productivity relation, the Omori-Utsu aftershock rate and B\r{a}th's relation in the spring-slider system, our comparison with the shuffled data or null model shows that these are simply consequences of the \emph{first}-order statistics, namely the avalanche size distribution and the inter-occurrence time distribution. 
This is because the null model, by construction, fully retains first-order statistics but disregards inherent correlations beyond a renewal process as captured by higher order statistics. Nevertheless, the null model is able to capture all the scaling features corresponding to the actual dynamical response. All these findings are robust over a wide range of relevant model parameters including the grain-level damping ratio, (sliding and/or rolling)  friction,  surface  roughness, and polydispersity (see Appendix). 
Moreover, our direct correlation analysis indicates a minimal presence of extremely short-ranged aftershocks at best in our system. 
As a result, we do not find any direct relevance of these correlations on the dynamics of avalanche sequences which are otherwise well-characterized by the productivity relationship, Omori-like temporal evolution, and B\r{a}th's relation.
This is at odds with natural seismicity, which exhibits strong (spatio-temporal) clustering effects commonly associated with aftershock dynamics.
Instead, our observations are consistent with the spring-slider experiment \cite{zadeh2019seismicity} where shuffling of the experimental data did not alter the productivity relationship, the Omori-Utsu relation, and B\r{a}th's relation either. 
This absence of aftershocks has also been observed in another experimental system~\cite{bares2018aftershock}, where individual acoustic events --- produced by a propagating tensile crack --- were fully described by the main (tectonic) seismicity relations without any correlations associated with the ordering of avalanches.
No evidence for aftershocks including no detectable temporal clustering was also found in fracturing experiments of intact rock samples~\cite{davidsen2017triggering,lennartz2014acceleration}.

With both processes governed by the regular stick-slip dynamics, our findings indicate an important difference between (tectonic) earthquakes and (plastic) slip avalanches in terms of the underlying relaxation mechanism. This includes differing origins of the observed scale-free behavior associated with the interevent time distributions in these two systems. 
In this context, Omori-type correlations associated with the earthquake dynamics largely determine the interevent time distributions as evidenced by the scaling relation $P(\tau)\propto \tau^{-(2-1/p)}$ with $p$ being the Omori exponent \cite{shcherbakov2005model}. 
However, due to the absence of pronounced correlations associated with (tectonic) aftershocks in our set-up, the proposed scaling relation is not applicable and we can in fact rule out such a (one-way) dependency between activity rates and $P(\tau)$.

It is noteworthy that we have only probed the slider motion in this study, not the internal dynamics of the rearranging grains, which lead to corresponding (acoustic) events during slip periods. Whether these internal events exhibit pronounced spatio-temporal correlations and aftershock sequences --- similar to some shearing  experiments~\cite{lherminier2019continuously} --- remains a challenge for the future.
Yet, with the granular substrate continually rearranging (and healing!) and in the absence of any memory effects (such as damage mechanism and/or frictional weakening), there is no obvious potential source of (spatio-)temporal clustering expected for (tectonic) aftershocks.

Within the framework of earthquake modeling, several treatments have been proposed in the literature which aim to incorporate physical relations leading to the generation of aftershock sequences (see \cite{de2016statistical} and references therein).
Common mechanisms such as the rate effects on solid friction \cite{scholz2002mechanics} and visco-elastic relaxations \cite{zhang2016power, baro2018universal} are conventionally implemented via basic phenomenological relations that typically involve a characteristic timescale in order to describe the dynamics of aftershocks.
As an essential feature associated with aftershock activities, (structural) heterogeneities have been also incorporated in several earthquake models involving inhomogeneous material parameters \cite{de2016statistical,pelletier2000spring}. 
These ingredients are at best mesoscopic and the underlying micro-structural processes that describe them are usually complex and not yet well understood.
Incorporating such microscopic features into spring-slider systems, either experimentally or in simulation, and/or fine-tuning the existing parameters may help recover the true dynamics of earthquakes.

\section{Conclusions}\label{sec:conclusions}

As a summary, we have identified common features as well as differences between the dynamics of ``micro" earthquakes, generated by a slowly-driven slider on a granular sublayer, and that of tectonic earthquakes at geological scales.
The former includes the Gutenberg-Richter relation, while the differences are largely related to the absence of aftershocks in our granular model. 
In particular, we established that the temporal correlations and clustering features associated with (tectonic) aftershocks are missing in the spring-slider setup and, therefore, the emerging Omori-Utsu relation, the aftershock productivity relation, and B{\aa}th's relation in the simulations have a fundamentally different origin from that of tectonic seismicity.
Specifically, a lack of temporal correlations in the experiment allows the derivation of the productivity relationship and (to some extent) B{\aa}th's relation entirely based on the Gutenberg-Richter statistics and without any further knowledge about the actual ordering of events in time.
This contrasts with the case of tectonic seismicity where the relevance of the productivity statistics is a direct consequence of aftershock dynamics.   
In the context of statistical seismology, to our knowledge, there is no established connection between the aftershocks productivity exponent $\alpha$ and the $b$-value describing the Gutenberg-Richter distribution, which indicates a main dissimilarity between natural earthquakes and slip avalanches in our slider system.

\clearpage

\appendix
\renewcommand{\thefigure}{A\arabic{figure}}
\setcounter{figure}{0}    

\renewcommand{\theequation}{A\arabic{equation}}
\setcounter{equation}{0}    

\section*{appendix}\label{AppA}

\subsection*{Robustness Analysis}
\newcommand{\ratioAppndx}{0.35}
Figure~\ref{fig:avlThresholds} shows the avalanche statistics for multiple thresholds $\dot{U}_c$, which is varied by more than an order of magnitude.
Avalanche size distributions in Fig.~\ref{fig:avlThresholds}(b) display robust power-law regimes with respect to variations in the chosen threshold, which only affects the lower cut-off corresponding to small avalanches.
In terms of the avalanche duration shown in Fig.~\ref{fig:avlThresholds}(a), higher thresholds result in shorter time scales (see Fig.~\ref{fig:EventDetails}) but leave the scaling relation between the avalanche size and duration basically unchanged.
We also checked that other relevant statistics (\emph{e.g.} productivity relation or aftershock rates) are robust; varying $\dot{U}_c$ led to rather small statistical fluctuations of the relevant scaling exponents (\emph{i.e.} productivity exponent $\alpha$ or $p$ exponent associated with the Omori-Utsu relation) around their mean values.

\begin{figure}[b]
     \begin{overpic}[width=\ratioAppndx\textwidth]{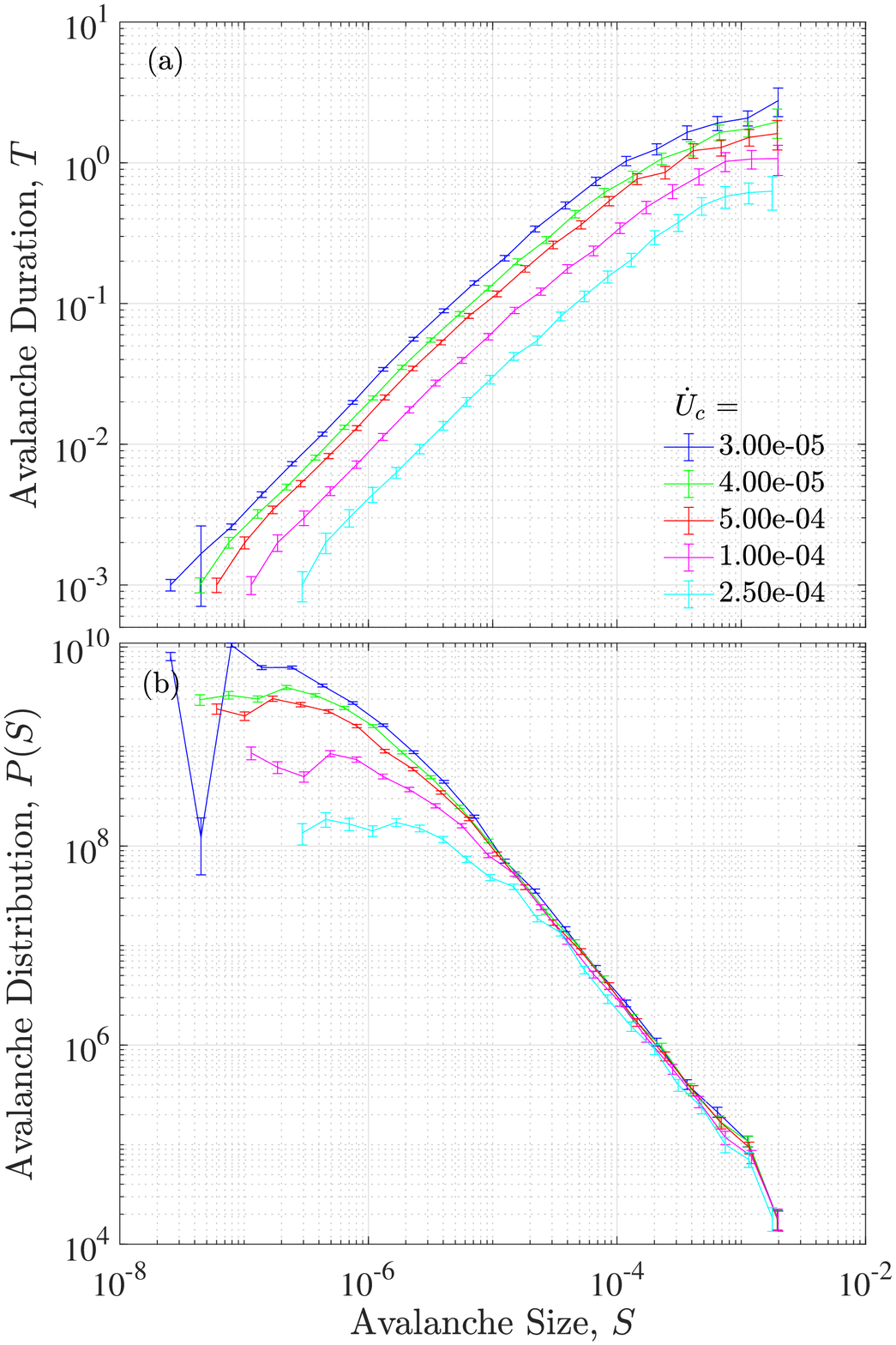}
         \centering
            \begin{tikzpicture}
              \coordinate (a) at (0,0); 
                \node[white] at (a) {\tiny.};  %
			\end{tikzpicture}
     \end{overpic}
         \caption{Avalanche statistics at multiple thresholds $\dot{U}_c$. \textbf{a)} Avalanche duration $T$ plotted against event size $S$ \textbf{b)} (non-normalized) avalanche size distribution $P(S)$.}
         \label{fig:avlThresholds}
\end{figure}

\begin{figure}[t]
     \begin{overpic}[width=\ratioAppndx\textwidth]{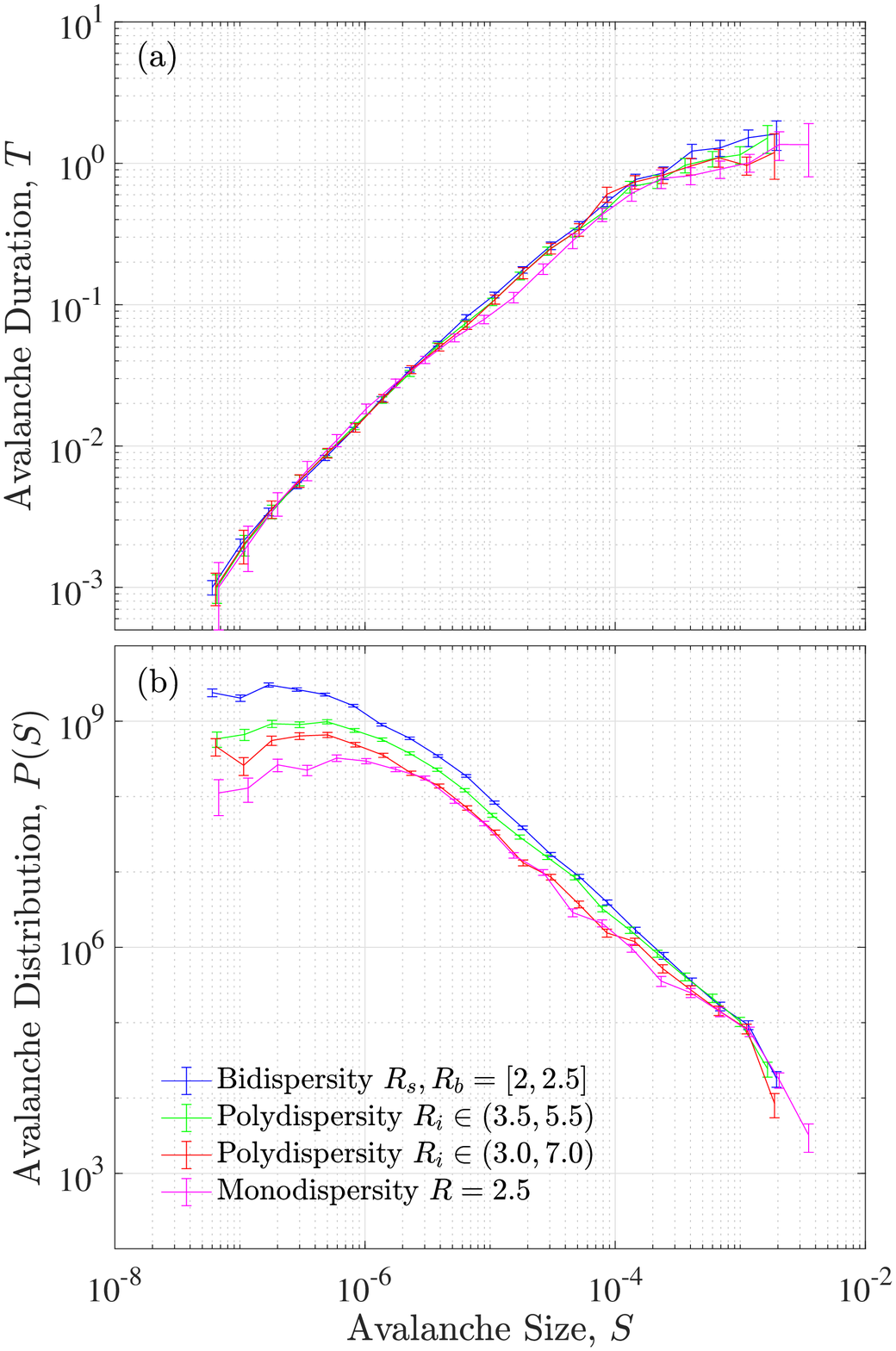}
         \centering
            \begin{tikzpicture}
              \coordinate (a) at (0,0); 
                \node[white] at (a) {\tiny.};  %
			\end{tikzpicture}
     \end{overpic}
         \caption{Avalanche statistics for different size distribution of substrate particles, sized denoted on the figure are in $mm$. \textbf{a)} Avalanche duration $T$ versus event size $S$ \textbf{b)} (non-normalized) avalanche size distribution $P(S)$.}
         \label{fig:Polydis}
\end{figure}

In Fig.~\ref{fig:Polydis}, we investigate potential effects of polydispersity on the statistics of avalanches. 
The particle size distributions include \emph{\romannum{1}}) monodisperse distribution with particle size of radius of $R = {2.5}$, \emph{\romannum{2}}) bidisperse (as in the experiment) with $R_b/R_s = 1.25$ and number ratio $N_b/N_s = 2.5$, and \emph{\romannum{3}}) two polydisperse cases, where particles are uniformly distributed between $R_i\in (R_\text{min},R_\text{max})$ with $R_i = R_\text{min}+i\times \Delta R$ denoting the radius for each species $i=0,1,2,...$. Specifically, we chose here $R_i\in (3.5,5.5)$ and $\Delta R=0.2$ for the first polydisperse case and $R_i\in (3.0,7.0)$ and $\Delta R=0.2$ for the second one.  The total number of particles, $N$, is roughly kept similar to the base case with $N = 7770$. 
Moreover, for each case the slider surface is kept the same. 
Fig.~\ref{fig:Polydis} indicates that polydisperse particle size distributions have no statistically significant effects on the avalanche statistics. 
In particular, the power law decay in the avalanche size distribution is found to be a robust feature across all cases.
Other statistics including the productivity relation, Omori-Utsu relationship, and B\r{a}th's relation are not significantly affected by variations in polydispersity (data not shown).

Furthermore, we explore the effect of changing the slider topology, i.e. the surface of the slider in contact with the granular matter. 
Specifically, we focus on reducing the depth of the grooves $r_i$ where the substrate disks stick, keeping all other parameters comparable to the regular case.
This reduction is such that the indentation left in the groove is a segment (measured in fraction) of the semicircle. 
The reduction in heights (measured upwards from the base of the slider) of the grooves, $h_{r}$, is chosen such that, $h_{r}/r_{i} = 0.6, 0.4$ and $0.2$; for example, the last case here is only the top segment of the semicircle.
The regular slider is the same as the base case with a complete arch of the semicircle with radius $r_{i} = 2.5~mm$, see Fig.~\ref{fig:SliderSystem}).

\begin{figure}[t]
     \begin{overpic}[width=\ratioAppndx\textwidth]{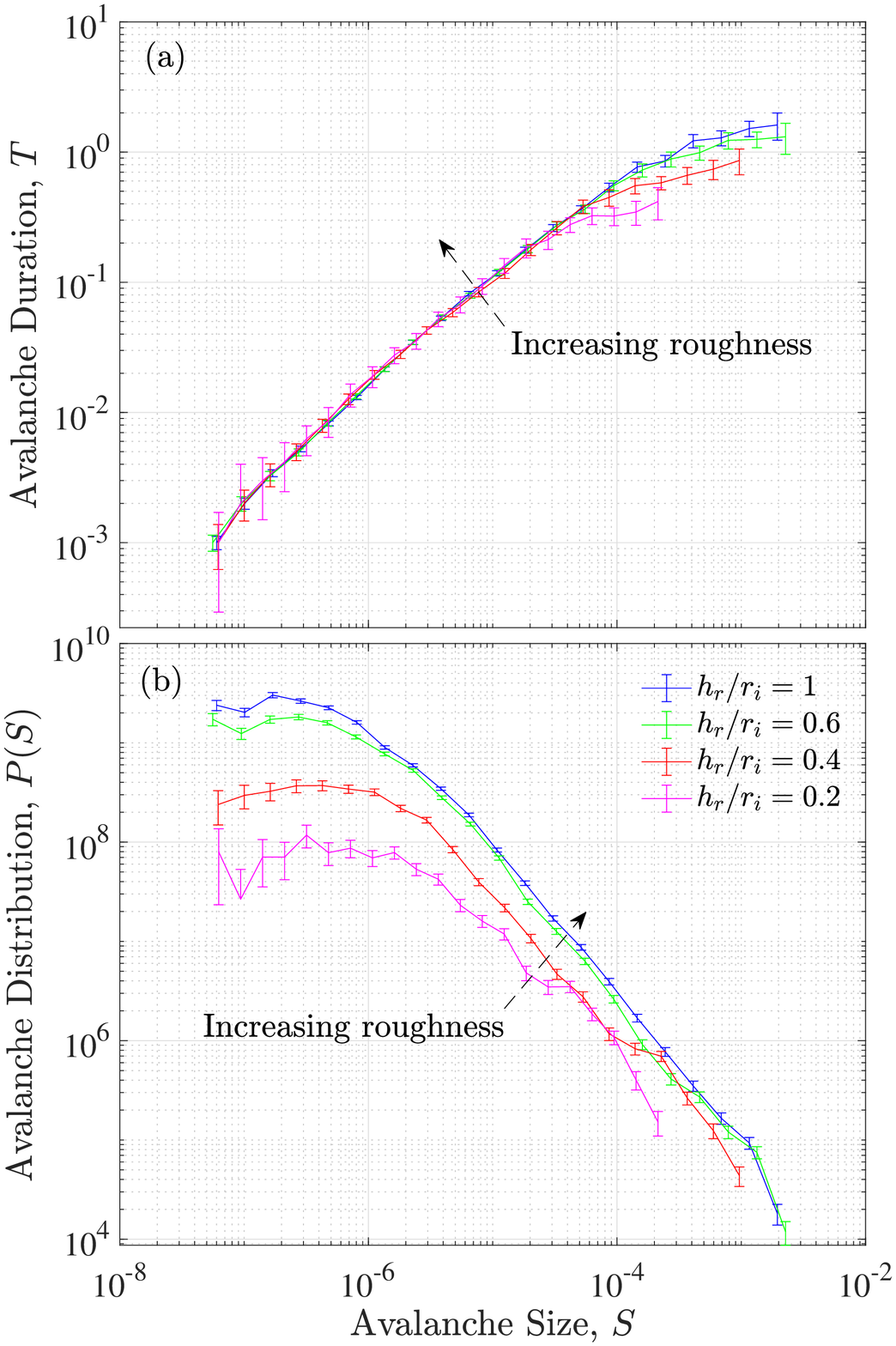}
         \centering
            \begin{tikzpicture}
              \coordinate (a) at (0,0); 
                \node[white] at (a) {\tiny.};  %
			\end{tikzpicture}
     \end{overpic}
         \caption{Avalanche statistics for different groove depth at the base of the slider, measured as a fraction, $h_{r}/r_{i}$. \textbf{a)} Avalanche duration $T$ versus event size $S$ \textbf{b)} (non-normalized) avalanche size distribution $P(S)$.}
         \label{fig:Slider}
\end{figure}

Figure~\ref{fig:Slider} shows the results of these different $h_{r}/r_{i}$ on avalanche duration and size. 
As the slider roughness is decreased (from $h_{r}/r_{i} = 1.0$ to $0.2$), the scaling range associated with the power-law regimes becomes narrower in $S$. Smoother sliders tend to have smaller upper cut-offs in the avalanche size distributions as well as larger lower cut-offs as indicated by an initial plateau regimes in Fig.~\ref{fig:Slider}(b). 
This feature, however, does not seem to have any significant effects on the scaling exponents $\gamma$ and $\beta$, particularly for $h_{r}/r_{i} = 1.0, 0.6, 0.4$.
These observations can be understood on the basis that sliders with smoother topology have a less pronounced stick-slip dynamics, similar to the case of higher pulling speeds.
Thus, we speculate that if one uses even smoother sliders, one might need to use slower pulling speeds in order to recover stick-slip dynamics and relevant scaling properties. 

\begin{figure}[b]
     \begin{overpic}[width=\ratioAppndx\textwidth]{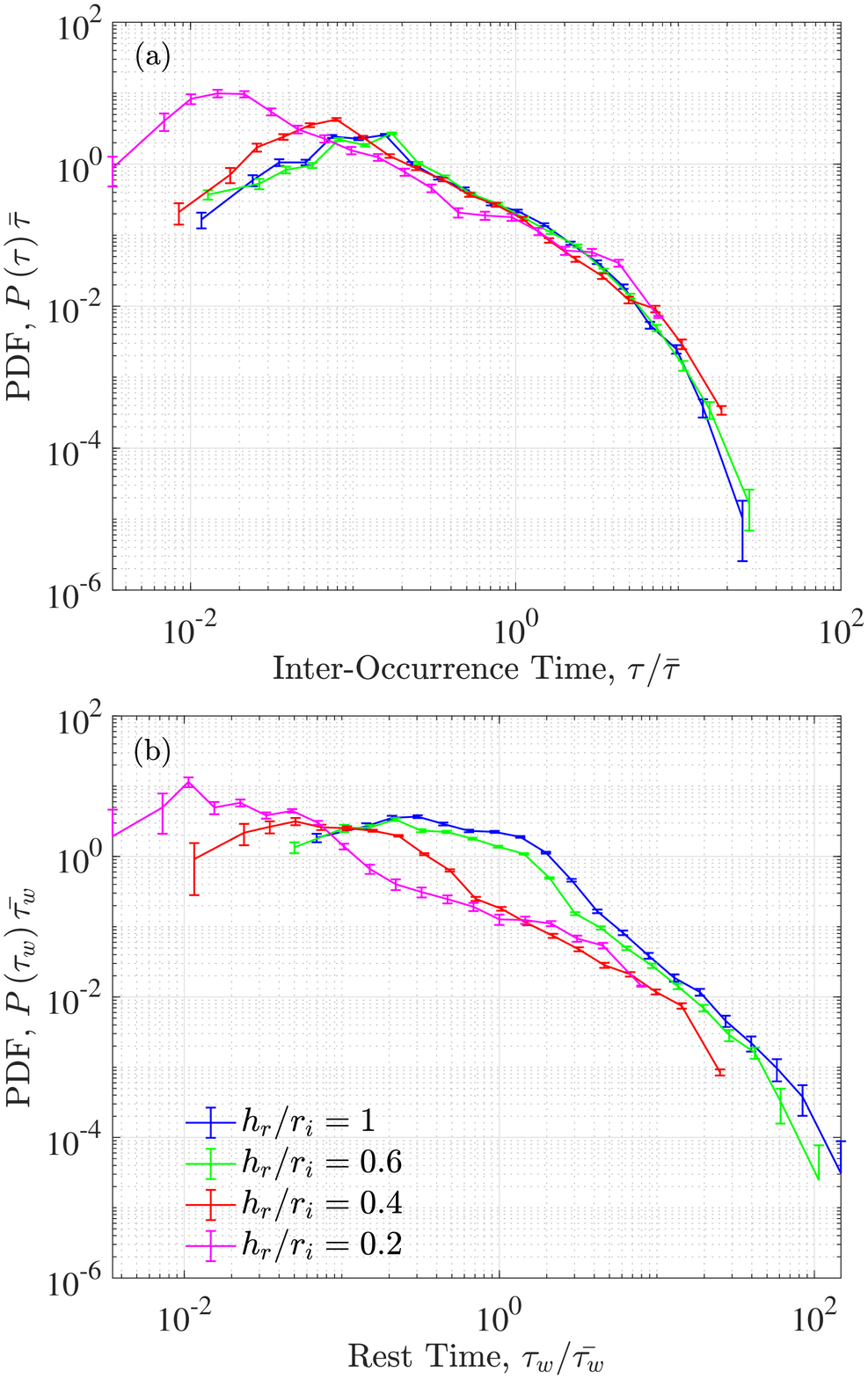}
         \centering
            \begin{tikzpicture}
              \coordinate (a) at (0,0); 
                \node[white] at (a) {\tiny.};
			\end{tikzpicture}
     \end{overpic}
         \caption{\textbf{a)} Inter-occurrence time distribution $\bar{\tau}P(\tau)$ and \textbf{b)} rest time distribution, $\bar{\tau}_wP(\tau_w)$, at different slider roughness, $h_r/r_i$. Here $\bar{\tau}$ and $\bar{\tau}_w$ indicate the  corresponding mean times. Energy threshold, $S_{c}$, chosen for these analyses is $10^{-6}$.}
         \label{fig:WaitingLaw}
\end{figure}

\begin{figure}[b]
     \begin{overpic}[width=\ratio\textwidth]{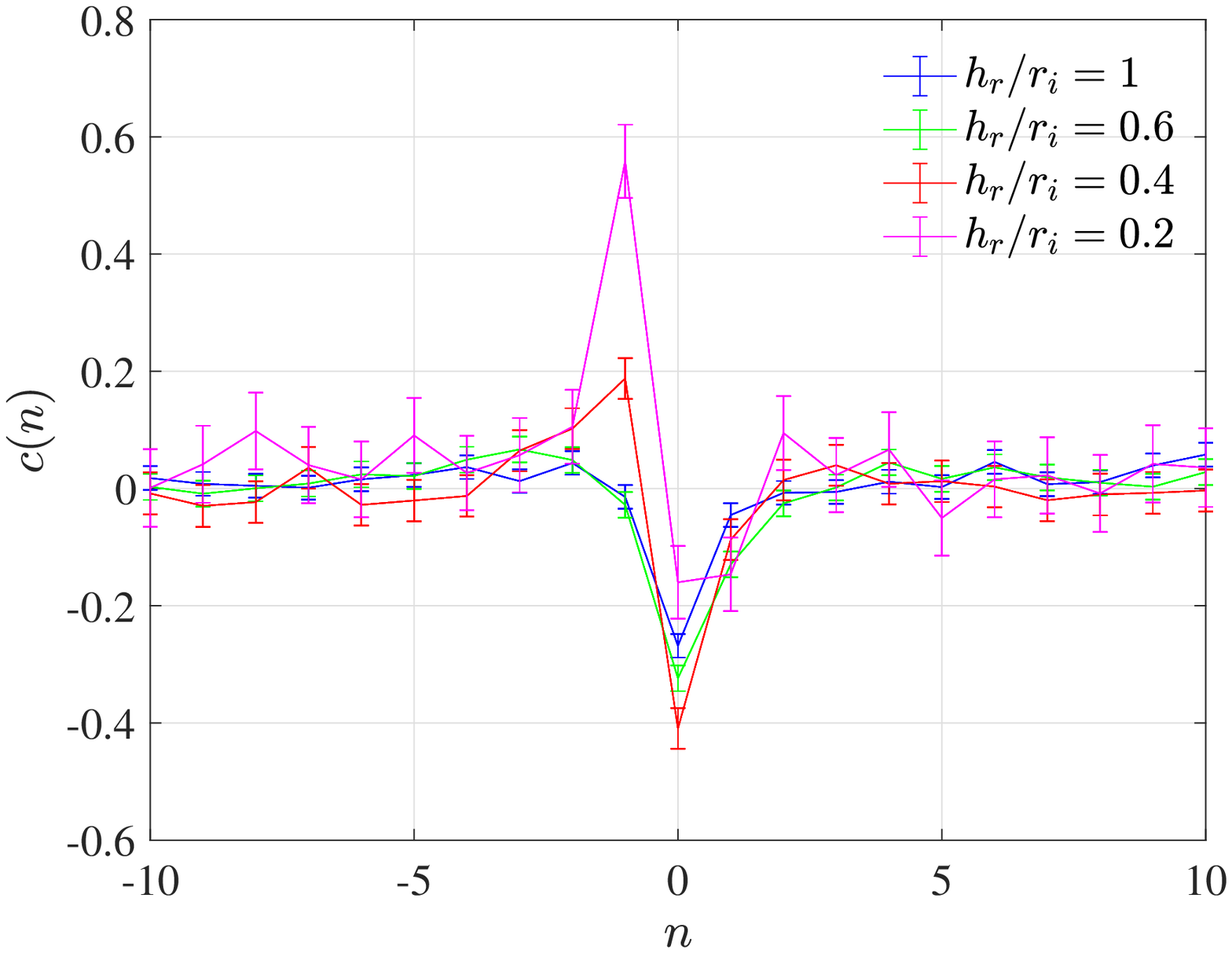}
         \centering
            \begin{tikzpicture}
              \coordinate (a) at (0,0); 
                \node[white] at (a) {\tiny.};  %
			\end{tikzpicture}
     \end{overpic}
         \caption{Cross correlation function $c(n)$ with index shift $n$ evaluated for different groove depth at the base of the slider, measured as a fraction, $h_{r}/r_{i}$. The (black) base line in indicates zero correlations. The error bars indicate one standard error.}
         \label{fig:crRoughness}
\end{figure}

Figure~\ref{fig:WaitingLaw} displays potential changes associated with distributions of inter-occurrence times $\tau$ and rest times $\tau_w$ for sliders with varying surface smoothness.
While $P(\tau)$'s are almost insensitive to variations in $h_r/r_i$, showing only an overall shift in both the lower and upper cut-offs, $P(\tau_w)$'s associated with the smoothest sliders seem to indicate a shallower power-law decay.
This might indeed be related to the observed change of temporal correlations shown in Fig.~\ref{fig:crRoughness} likely due to the emerging quasi-periodic sliding regime.

Figure~\ref{fig:crRoughness} examines the dependence of temporal correlations (cf. Fig.~\ref{fig:crltnFunction}) on the slider smoothness which seems to control the existence and height of the positive peak at $n=-1$.
A positive correlation (at negative $n$ values) implies that big events are, on average, preceded by long rest times as expected for quasi-periodic behavior. These positive correlations become quite insignificant at the two roughest slider surfaces, i.e. $h_r/r_i=0.6, 1.0$.
In contrast, the negative peak at $n= 0$ is a rather robust feature with respect to variations in $h_r/r_i$. 

\begin{figure}
     \begin{overpic}[width=\ratioAppndx\textwidth]{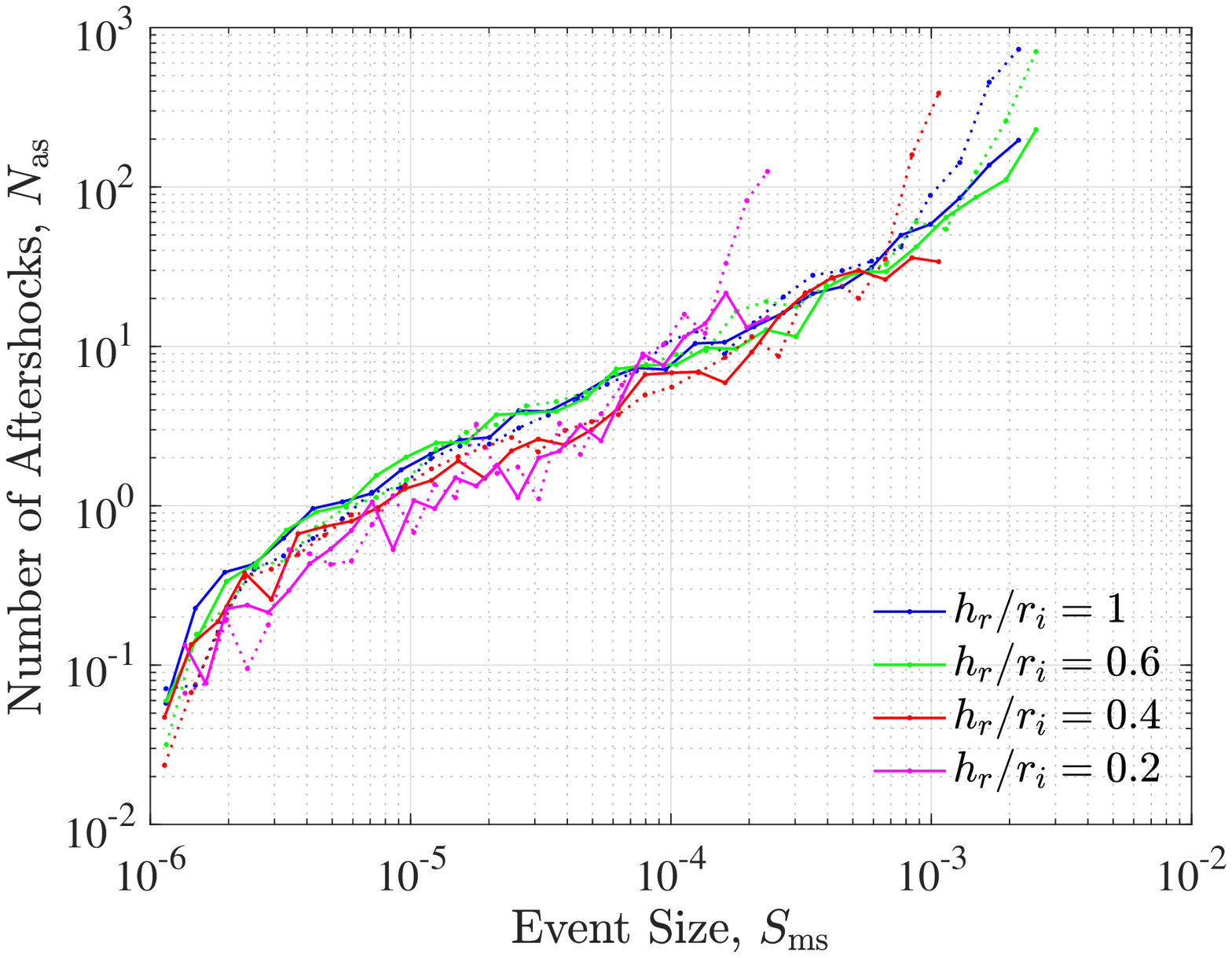}
         \centering
            \begin{tikzpicture}
              \coordinate (a) at (0,0); 
                \node[white] at (a) {\tiny.};
			\end{tikzpicture}
     \end{overpic}
         \caption{Number of aftershocks $N_{as}$ for a mainshock size $S_{ms}$ shown for different sliders. The solid lines show the unshuffled cases and dotted lines show the shuffled cases.}
         \label{fig:SliderProductivity}
\end{figure}

\begin{figure}[b]
     \begin{overpic}[width=0.5\textwidth]{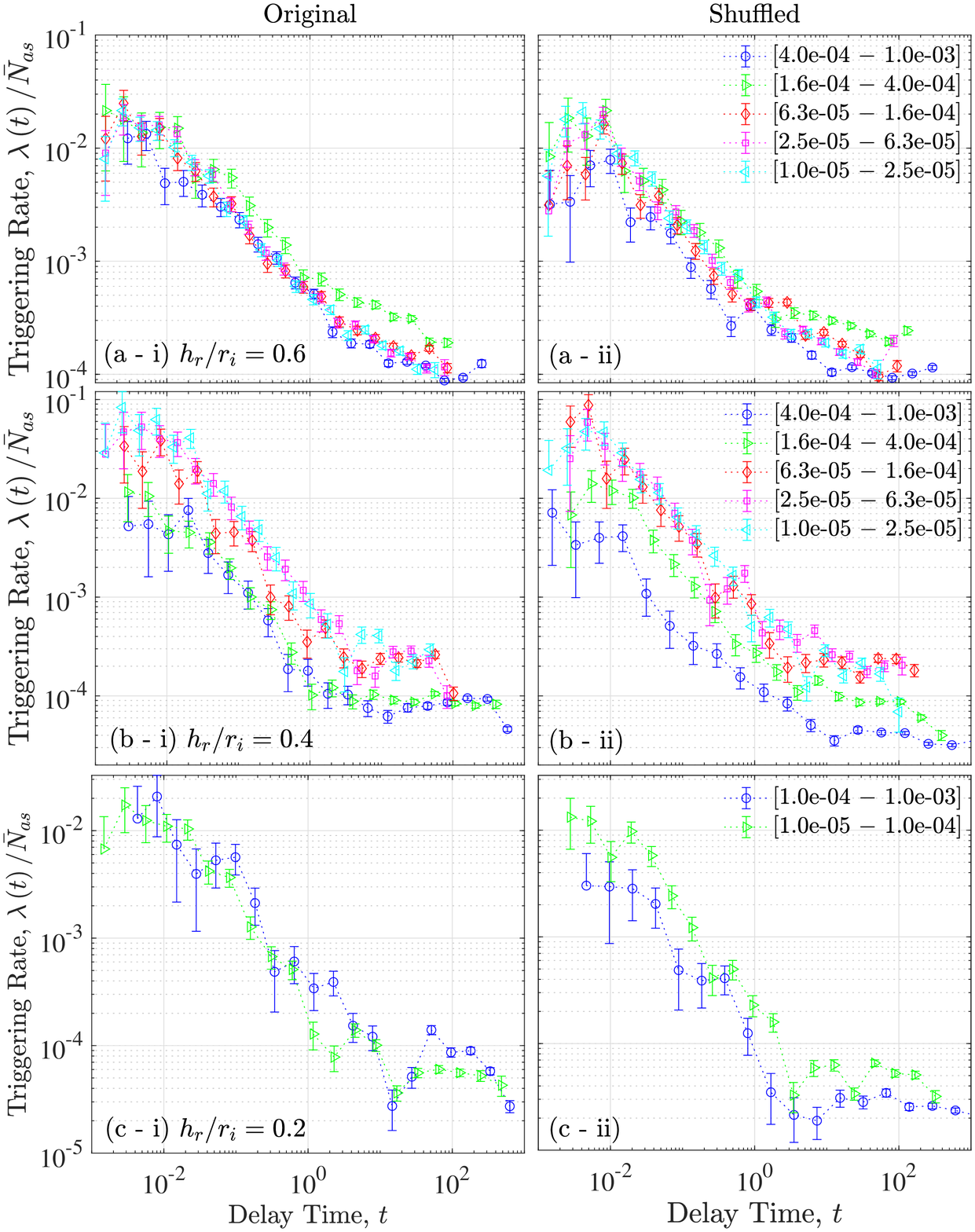}
         \centering
            \begin{tikzpicture}
              \coordinate (a) at (0,0); 
                \node[white] at (a) {\tiny.};
			\end{tikzpicture}
     \end{overpic}
         \caption{Triggering rates, $\lambda\left(t\right)$, normalized by average number of after-shocks $\bar{N}_{as}$ for different sliders topologies. Rows labeled (a), (b) and (c) show results for sliders cases $h_{r}/r_{i} = 0.6, 0.4$ and $0.2$ respectively. Left column sub-labeled (~-~i) are unshuffled (original) cases and right column sub-labeled (~-~ii) are results for shuffled cases.}
         \label{fig:SliderOmori}
\end{figure}

\begin{figure}
     \begin{overpic}[width=\ratioAppndx\textwidth]{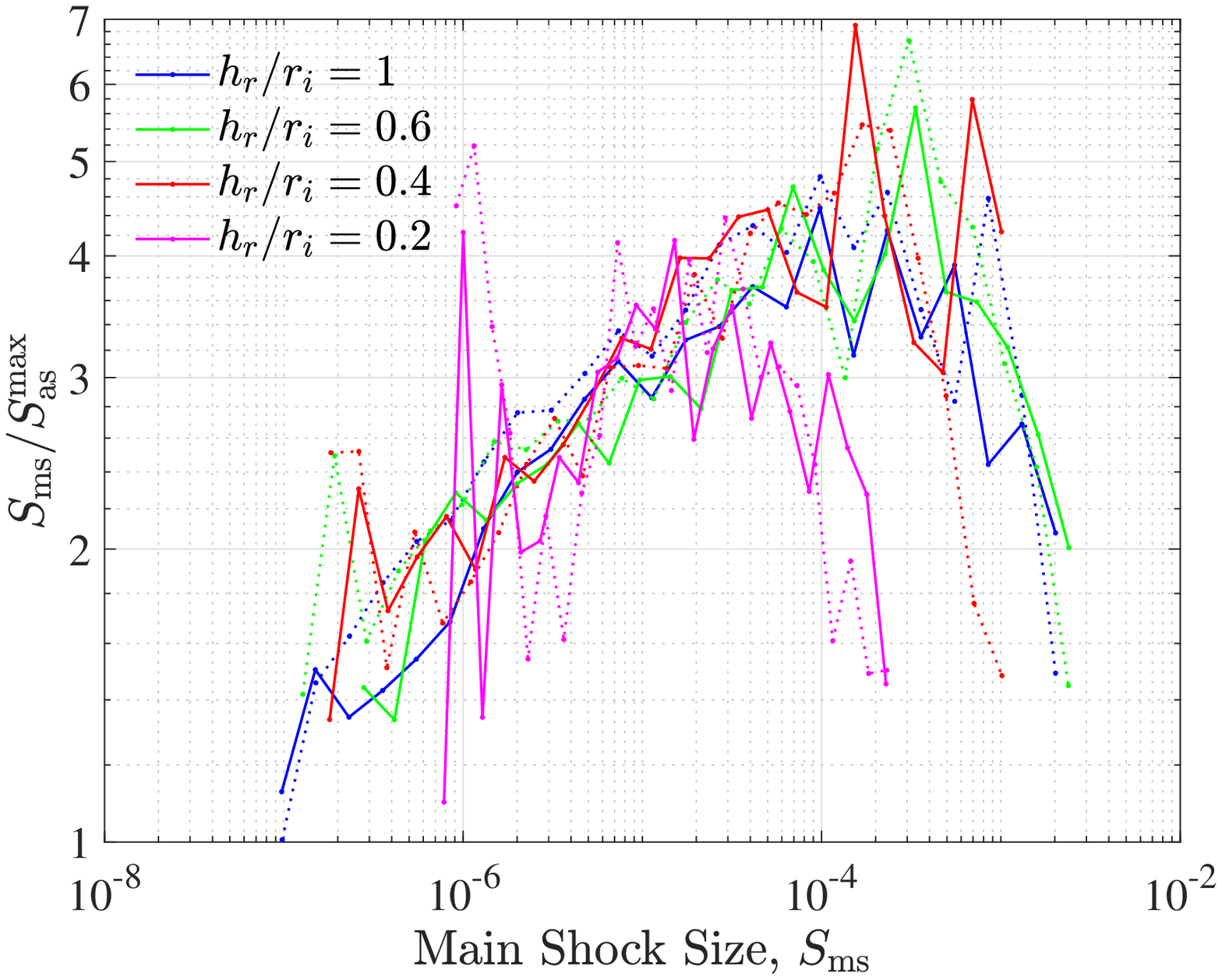}
         \centering
            \begin{tikzpicture}
              \coordinate (a) at (0,0); 
                \node[white] at (a) {\tiny.};
			\end{tikzpicture}
     \end{overpic}
         \caption{Mainshock energies normalized by the maximum aftershock energies in a given mainshock-aftershock sequences for different sliders. The solid lines show the unshuffled cases and dotted lines show the shuffled cases.}
         \label{fig:SliderBath}
\end{figure}

Figure~\ref{fig:SliderProductivity} illustrates the robustness of the productivity relationship against variations in $h_r/r_i$. 
Likewise, temporal aftershock rates shown in Fig.~\ref{fig:SliderOmori} indicate a robust scaling regime with respect to the changes of the slider surface.
We also show robust features associated with the B\r{a}th's relation in Fig.~\ref{fig:SliderBath}.

\subsection*{Revisiting B\r{a}th's Relation and Productivity Relationship}
We provide a formal theoretical framework to establish Bath’s relation. Let $P(m)=\lambda e^{-\lambda m}$ and $F(m)=1-e^{-\lambda m}$ be the probability distribution and the accumulated distribution function of magnitude $m$, respectively.
Here $\lambda$ controls the rate of exponential decay.
For a given main shock-aftershock sequence 
\begin{equation}\label{eq:sequence}
\{m_\text{ms};~m_1<m_\text{ms},~...,~m_N<m_\text{ms},~m_{N+1}>m_\text{ms}\},
\end{equation}
let the largest aftershock magnitude be $M=\text{max}\{m_1,~...,~m_N\}$.
Assuming that events are independent, it follows that 
\begin{eqnarray}\label{eq:extremeVals}
F(M |m_\text{ms},N)&=&(1-e^{-\lambda M})^N, \nonumber \\ P(M|m_\text{ms},N)&=& \partial_M F(M|m_\text{ms},N) \nonumber \\ 
&=& N\lambda e^{-\lambda M} (1-e^{-\lambda M})^{N-1}.
\end{eqnarray}
We note that, for large $N$ and $m_\text{ms}\rightarrow \infty$, the (cumulative) distribution for the \emph{scaled} magnitude $z=(M-a)/b$ will asymptotically converge to the Gumbel distribution $F(z)=e^{-e^{-z}}$.
Here $a=1/\lambda$ and $b=\text{log}_{e}(N)/\lambda$.
Given that $\langle z\rangle\simeq 0.5772$, the conditional mean is
\begin{equation}\label{eq:asymMean}
\langle M|m_\text{ms}\rightarrow \infty,~N\rangle \simeq \{\langle z\rangle+\text{log}_{e}(N)\}/\lambda,
\end{equation}
 which grows logarithmically with $N$.

One could subsequently derive the B\r{a}th's relation by performing the following summation 
\begin{equation}\label{eq:condtnlMean}
\langle M|m_\text{ms}\rangle =\sum_{N=1}^{\infty}\langle M|m_\text{ms},N\rangle ~P(N|m_\text{ms}),
\end{equation}
where 
\begin{equation}\label{eq:seqPr}
P(N|m_\text{ms})=(1-e^{-\lambda m_\text{ms}})^N~e^{-\lambda m_\text{ms}},
\end{equation}
 is the probability of having the main shock-aftershock sequence in Eq.~\ref{eq:sequence}.
 We performed the numerical integration using Eqs.~\ref{eq:extremeVals}, \ref{eq:condtnlMean}, and \ref{eq:seqPr} and obtained $\langle M|m_\text{ms}\rangle=m_\text{ms}-1/\lambda$. 
 Rewriting this relation using the energy scale, e.g. $m=\text{log}_{10}S$, we have $S_\text{ms}/S_\text{as}^\text{max}=e^{1/(\beta-1)}$ with $\beta$ denoting the avalanche size exponent.

In terms of the productivity relation, we may also evaluate the mean aftershock number conditioned on the main shock magnitude as
\begin{eqnarray}
\langle N|m_\text{ms}\rangle &=& \sum_{N=0}^\infty N~P(N|m_\text{ms}) \nonumber \\
&=& (1-e^{-\lambda m_\text{ms}})~e^{\lambda m_\text{ms}}.
\end{eqnarray}
For large $m_\text{ms}$, it follows that $\langle N|m_\text{ms}\rangle \propto e^{\lambda m_\text{ms}}$. 
Therefore, $\bar{N}_\text{as}\propto S_\text{ms}^{\beta-1}$.

\acknowledgements
JD was financially supported by the Natural Sciences and Engineering Research Council of Canada (NSERC).

\clearpage
\bibliography{ref}

\end{document}